\documentclass[letter,twocolumn,showpacs,aps,prd,preprintnumbers,superscriptaddress]{revtex4}
\usepackage{graphicx}
\usepackage{dcolumn}
\usepackage{amsmath}
\usepackage{epsfig}
\usepackage{color}

\usepackage{relsize}
\RequirePackage{xspace}

\def\pep2{PEP-II}
\def\babar{\mbox{\slshape B\kern-0.1em{\smaller A}\kern-0.1em
    B\kern-0.1em{\smaller A\kern-0.2em R}}}
\def\cm   {\ensuremath{{\rm \,cm}}\xspace}
\def\Y#1S{\ensuremath{\Upsilon{(#1S)}}\xspace}
\def\FourS {\Y4S}
\def\KL    {\ensuremath{K^0_{\scriptscriptstyle L}}\xspace}
\def\ep         {\ensuremath{e^+}\xspace}
\def\epem       {\ensuremath{e^+e^-}\xspace}
\def\mup        {\ensuremath{\mu^+}\xspace}
\def\mun        {\ensuremath{\mu^-}\xspace}

\def\piz   {\ensuremath{\pi^0}\xspace}
\def\pip   {\ensuremath{\pi^+}\xspace}
\def\pim   {\ensuremath{\pi^-}\xspace}
\def\Kp    {\ensuremath{K^+}\xspace}
\def\Km    {\ensuremath{K^-}\xspace}
\def\Dbar    {\kern 0.2em\overline{\kern -0.2em D}{}\xspace}
\def\Db      {\ensuremath{\Dbar}\xspace}
\def\Dz      {\ensuremath{D^0}\xspace}
\def\Dzb     {\ensuremath{\Dbar^0}\xspace}
\def\DzDzb   {\ensuremath{\Dz {\kern -0.16em \Dzb}}\xspace}
\def\Dp      {\ensuremath{D^+}\xspace}
\def\Dm      {\ensuremath{D^-}\xspace}
\def\DpDm    {\ensuremath{\Dp {\kern -0.16em \Dm}}\xspace}
\def\Dstar   {\ensuremath{D^*}\xspace}
\def\jpsi     {\ensuremath{{J\mskip -3mu/\mskip -2mu\psi\mskip 2mu}}\xspace}

\def\etac     {\ensuremath{\eta_c}\xspace}
\def\chiczero {\ensuremath{\chi_{c0}}\xspace}
\def\chicone  {\ensuremath{\chi_{c1}}\xspace}
\def\chictwo  {\ensuremath{\chi_{c2}}\xspace}
\newcommand{\tev}{\ensuremath{\mathrm{\,Te\kern -0.1em V}}\xspace}
\newcommand{\gev}{\ensuremath{\mathrm{\,Ge\kern -0.1em V}}\xspace}
\newcommand{\mev}{\ensuremath{\mathrm{\,Me\kern -0.1em V}}\xspace}
\newcommand{\kev}{\ensuremath{\mathrm{\,ke\kern -0.1em V}}\xspace}
\newcommand{\ev}{\ensuremath{\mathrm{\,e\kern -0.1em V}}\xspace}
\newcommand{\gevc}{\ensuremath{{\mathrm{\,Ge\kern -0.1em V\!/}c}}\xspace}
\newcommand{\mevc}{\ensuremath{{\mathrm{\,Me\kern -0.1em V\!/}c}}\xspace}
\newcommand{\gevcc}{\ensuremath{{\mathrm{\,Ge\kern -0.1em V\!/}c^2}}\xspace}
\newcommand{\mevcc}{\ensuremath{{\mathrm{\,Me\kern -0.1em V\!/}c^2}}\xspace}
\def\fb   {\ensuremath{\mbox{\,fb}}\xspace}
\def\invfb   {\ensuremath{\mbox{\,fb}^{-1}}\xspace}
\newcommand{\chisq}{\ensuremath{\chi^2}\xspace}
\newcommand{\epjBase}        {Eur.\ Phys.\ Jour.\xspace}
\newcommand{\jprlBase}       {Phys.\ Rev.\ Lett.\xspace}
\newcommand{\jprBase}        {Phys.\ Rev.\xspace}
\newcommand{\jplBase}        {Phys.\ Lett.\xspace}
\newcommand{\nimBaseA}       {Nucl.\ Instrum.\ Methods Phys.\ Res., Sect.\ A\xspace}

\newcommand{\npBase}         {Nucl.\ Phys.\xspace}

\newcommand{\epjc}      [1]  {\epjBase\ C~{\bf #1}}

\newcommand{\ijmpa}     [1]  {{Int.\ J.\ Mod.\ Phys.\ {\bf A{\bf #1}}}}

\newcommand{\nima}      [1]  {\nimBaseA~{\bf #1}}

\newcommand{\np}        [1]  {\npBase\ {\bf #1}}
\newcommand{\npb}       [1]  {\npBase\ B~{\bf #1}}

\newcommand{\plb}       [1]  {\jplBase\ B~{\bf #1}}
\newcommand{\prep}      [1]  {{Phys.\ Rep.\ {\bf #1}}}
\newcommand{\jprl}      [1]  {\jprlBase\ {\bf #1}}

\newcommand{\jprd}      [1]  {\jprBase\ D~{\bf #1}}

\newcommand{\BABARPubYear}    {09}
\newcommand{\BABARPubNumber}  {031}

\newcommand{\SLACPubNumber} {13939}
\newcommand{\LANLNumber} {1002.0281}

\def\figurebox#1#2#3{%
    \def\arg{#3}%
    \ifx\arg\empty
    {\hfill\vbox{\hsize#2\hrule\hbox to #2{\vrule\hfill\vbox to #1{\hsize#2\vfill}\vrule}\hrule}\hfill}%
    \else
    {\hfill\epsfbox{#3}\hfill}%
    \fi}

\begin{document}

\preprint{\babar-PUB-\BABARPubYear/\BABARPubNumber}
\preprint{SLAC-PUB-\SLACPubNumber}
\preprint{arXiv:\LANLNumber\ [hep-ex]}

\title{\large \bf Observation of the {\boldmath $\chi_{c2}(2P)$} meson in the reaction {\boldmath $\gamma\gamma \to D\Db$} at \babar}

\author{B.~Aubert}
\author{Y.~Karyotakis}
\author{J.~P.~Lees}
\author{V.~Poireau}
\author{E.~Prencipe}
\author{X.~Prudent}
\author{V.~Tisserand}
\affiliation{Laboratoire d'Annecy-le-Vieux de Physique des Particules (LAPP), Universit\'e de Savoie, CNRS/IN2P3,  F-74941 Annecy-Le-Vieux, France}
\author{J.~Garra~Tico}
\author{E.~Grauges}
\affiliation{Universitat de Barcelona, Facultat de Fisica, Departament ECM, E-08028 Barcelona, Spain }
\author{M.~Martinelli$^{ab}$}
\author{A.~Palano$^{ab}$ }
\author{M.~Pappagallo$^{ab}$ }
\affiliation{INFN Sezione di Bari$^{a}$; Dipartimento di Fisica, Universit\`a di Bari$^{b}$, I-70126 Bari, Italy }
% ...
\collaboration{The \babar\ Collaboration}
\noaffiliation

\date{\today}

\begin{abstract}
A search for the $Z(3930)$ resonance in $\gamma\gamma$ production of the $D\Db$ system has been performed using a data sample corresponding to an integrated luminosity of $384\invfb$ recorded by the \babar~experiment at the PEP-II asymmetric-energy electron-positron collider. The $D\Db$ invariant mass distribution shows clear evidence of the $Z(3930)$ state with a significance of $5.8 \sigma$. We determine mass and width values of $(3926.7 \pm 2.7 \pm 1.1)\mevcc$ and $(21.3 \pm 6.8 \pm 3.6)\mev$, respectively. A decay angular analysis provides evidence that the $Z(3930)$ is a tensor state with positive parity and $C$-parity ($J^{PC} = 2^{++}$); therefore we identify the $Z(3930)$ state as the $\chictwo(2P)$ meson. The value of the partial width $\Gamma_{\gamma\gamma}\times {\cal B}(Z(3930)\to D\Db)$ is found to be $(0.24 \pm 0.05 \pm 0.04)\kev$.  
\end{abstract}

\pacs{13.66.Bc, 14.40.Gx, 13.25.Gv}
\maketitle

\section{Introduction}
\begin{figure}
\includegraphics[width=0.3\textwidth]{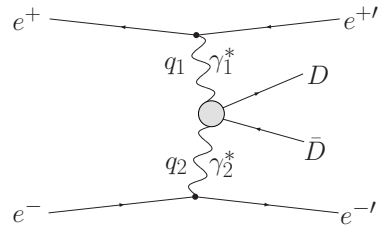}
\caption{Two-photon production of the $D\Db$ system.}
\label{fig:ggfeyn}
\end{figure}

Interest in the field of charmonium spectroscopy has been renewed with the recent discovery of numerous charmonium and charmonium-like states~\cite{X3872a, X3872b, X3872c, X3872d, X3940, Y3940a, Y3940b, Y4260a, Y4260b, Y4360a, Y4360b}. However very little is known about the first radially excited $\chi_{cJ}(2P)$ states which are expected to exist in the mass region from $3.9$ to $4.0\gevcc$, just above the $D\Db$ threshold~\cite{Go85}. The Belle collaboration has observed the $Z(3930)$ state in $\gamma\gamma$ production of the $D\Db$ system~\cite{Ue06}, and this is considered a strong candidate for the $\chictwo(2P)$ state; indeed it is so-labelled in Ref.~\cite{Pd08}. The Belle analysis obtained a mass of $m = (3929 \pm 5 \pm 2)\mevcc$ and a total width of $\Gamma = (29 \pm 10 \pm 2)\mev$, with quantum numbers $J^{PC} = 2^{++}$ preferred. The partial width $\Gamma_{\gamma\gamma}\times {\cal B}(Z(3930)\to D\Db)$ was determined as $(0.18 \pm 0.05 \pm 0.03)\kev$, where $\Gamma_{\gamma\gamma}$ is the radiative width of the $Z(3930)$ state, under the assumption that $J = 2$. The observation of this state has not been confirmed so far~\cite{Pd08}.\\
\indent In this paper the process $\gamma\gamma \to D\Db$, illustrated by the Feynman diagram shown in Fig.~\ref{fig:ggfeyn}, is studied in a search for the $Z(3930)$ state. In Fig.~\ref{fig:ggfeyn}, the initial state positron, $\ep$ (electron, $e^{-}$), emits the virtual photon $\gamma_{1}^{\ast}$ ($\gamma_{2}^{\ast}$), yielding the final state positron, $e^{+\prime}$ (electron, $e^{-\prime}$); the momentum transfer to $\gamma_{1}^{\ast}$ ($\gamma_{2}^{\ast}$) is $q_{1}$ ($q_{2}$). The virtual photons interact to produce the $D\Db$ final state. When the $e^{+\prime}$ and the $e^{-\prime}$ are emitted along the beam directions the values of $q_{1}^{2}$ and $q_{2}^{2}$ are predominantly close to zero, and the two photons can be considered to be quasi-real. Since in this case neither the $e^{+\prime}$ nor the $e^{-\prime}$ are detected, the analysis is termed {\it untagged}.\\
\indent The data sample used in this analysis corresponds to an integrated luminosity of $384\invfb$ recorded at the $\FourS$ resonance ($10.58\gev$) and at a center of mass (c.m.)~energy of $10.54\gev$ by the \babar~detector at the PEP-II asymmetric-energy $\epem$ collider.\\
\indent The \babar~detector is described briefly in Sec.~\ref{sec_babdet}, and the principal criteria used in the selection of candidate two-photon-interaction events are discussed in Sec.~\ref{sec_ggsel}. The reconstruction of $D\Db$ pair events is presented in Sec.~\ref{sec_reco}, and the relevant Monte Carlo simulations are detailed in Sec.~\ref{sec_mcstud}. The purity and reconstruction efficiency of the $\gamma\gamma \to D\Db$ event sample are considered in Secs.~\ref{sec_puri} and~\ref{sec_effi}, respectively, and the $D\Db$ signal yield and invariant mass resolution are presented in Sec.~\ref{sec_fit}; the mass and total width for the $Z(3930)$ state are obtained from a fit to the $D\Db$ invariant mass distribution. The angular distribution in the $D\Db$ rest frame for the $Z(3930)$ mass region is studied in Sec.~\ref{sec_ang}, and the implications for the spin of the $Z(3930)$ state are discussed. In Sec.~\ref{sec_rawi} the partial radiative width of the $Z(3930)$ state is extracted. Sources of systematic uncertainty are detailed in Sec.~\ref{sec_syst}, and the results of the analysis are summarized in Sec.~\ref{sec_summ}.

\section{The \babar~Detector}
\label{sec_babdet}

The \babar~detector is described in detail elsewhere~\cite{Au02}. Charged particles are detected, and their momenta measured, with a combination of five layers of double-sided silicon microstrip detectors (SVT) and a 40-layer cylindrical drift chamber (DCH), both coaxial with the cryostat of a superconducting solenoidal magnet which produces a magnetic field of $1.5~\mathrm{T}$. Charged particle identification is achieved by measurements of the energy loss $dE/dx$ in the tracking devices and by means of an internally reflecting, ring-imaging Cherenkov detector (DIRC). Photons and electrons are detected and their energies measured with a CsI(Tl) electromagnetic calorimeter (EMC), covering $90\%$ of the $4\pi$ solid angle in the $\Upsilon(4S)$ rest frame. The instrumented flux return of the magnetic field is used to identify muons and \KL. 

\section{Selection of Two-photon-interaction events}
\label{sec_ggsel}
\begin{figure*}
\includegraphics[width=0.3\textwidth]{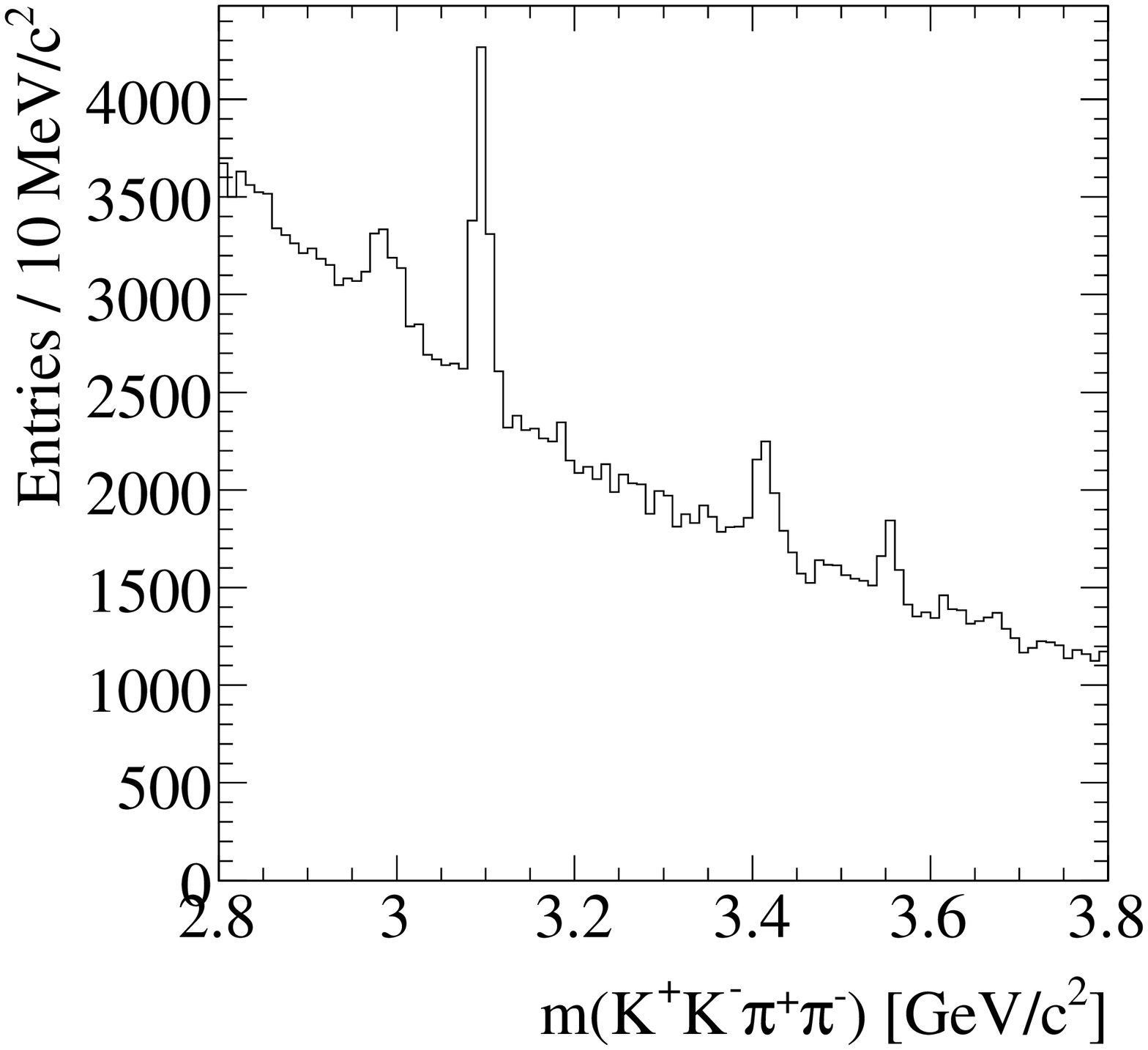}
\includegraphics[width=0.3\textwidth]{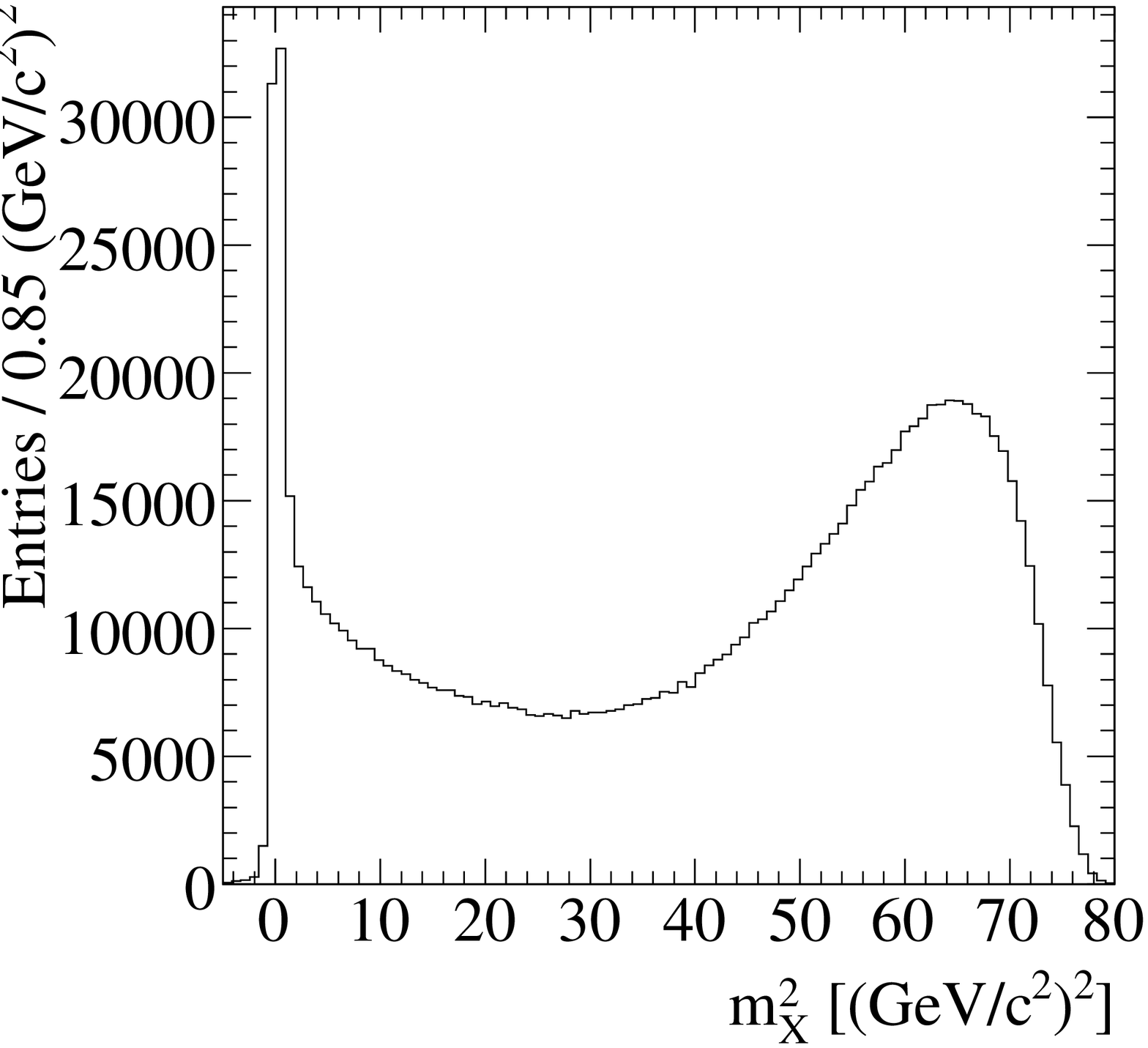}\\
\includegraphics[width=0.3\textwidth]{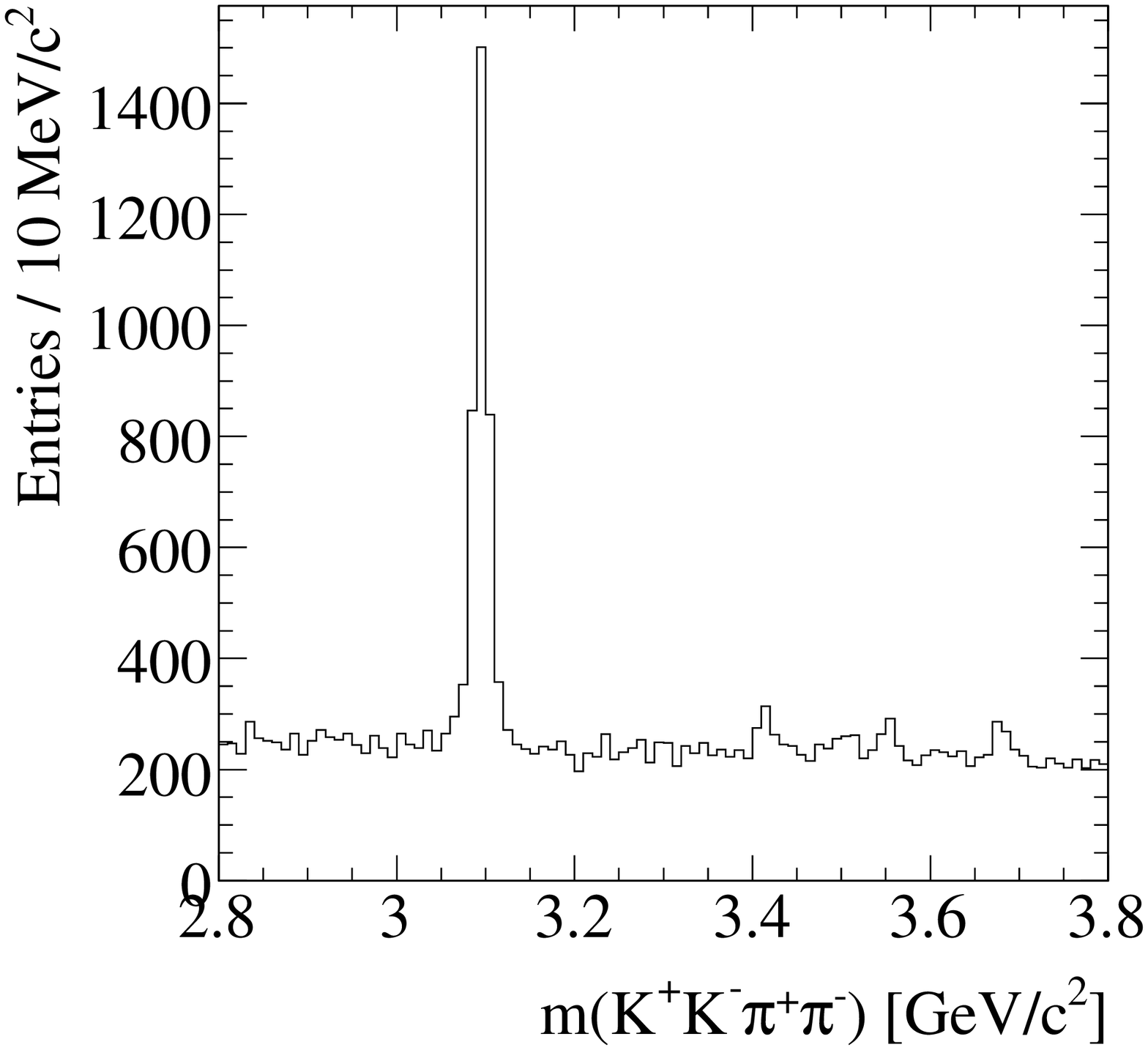}
\includegraphics[width=0.3\textwidth]{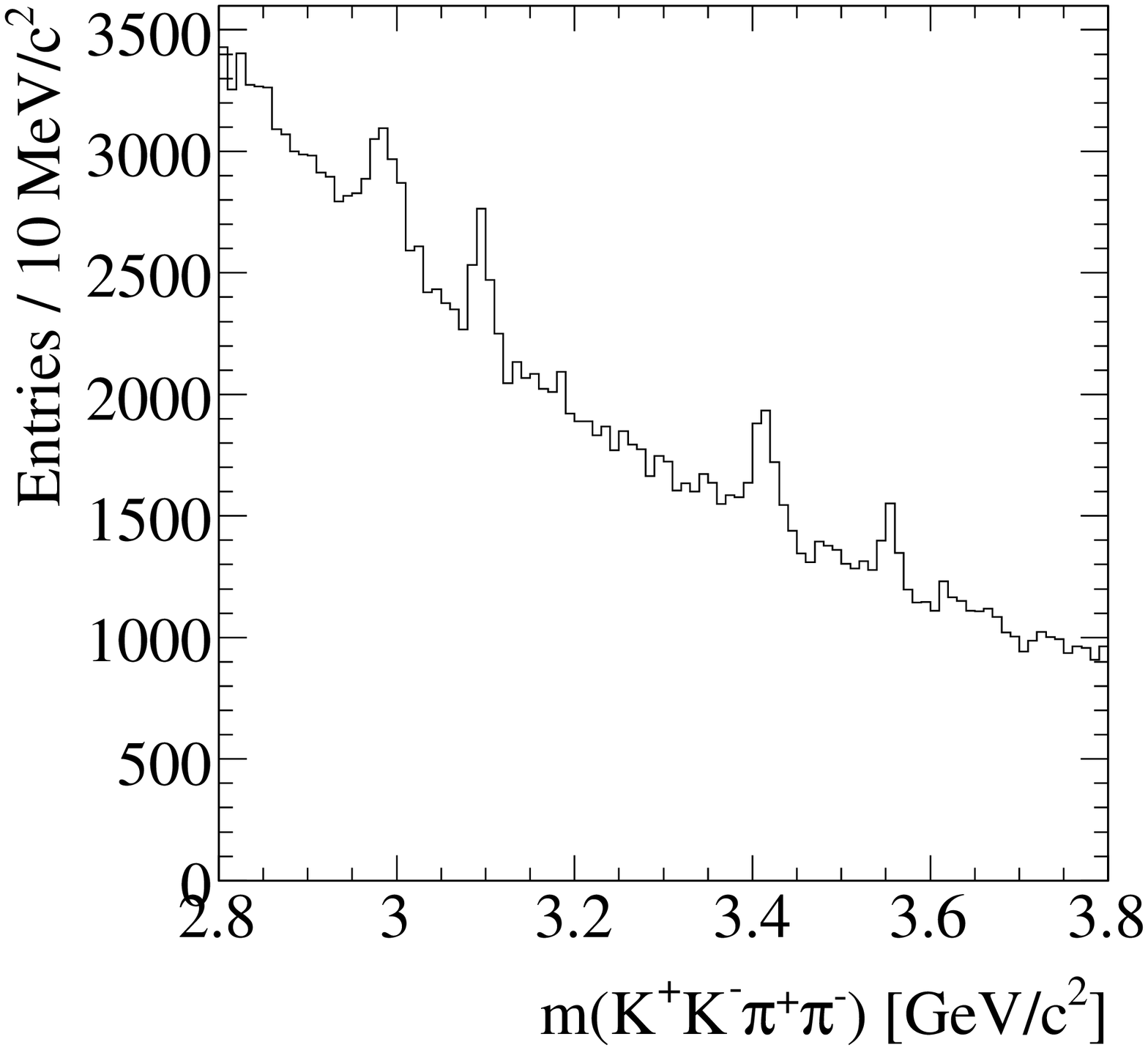}
\caption{(a) $\Kp\Km\pip\pim$ mass distribution for all events without any requirement on $m_{X}^{2}$; (b) corresponding $m_{X}^{2}$ distribution;  (c) $m(\Kp\Km\pip\pim)$ with the requirement $m_{X}^{2} < 10~(\!\gevcc)^{2}$; (d) $m(\Kp\Km\pip\pim)$ with the requirement $m_{X}^{2} > 10~(\!\gevcc)^{2}$.}
\begin{picture}(0,0)
\put(-1.5,10.5){(a)}
\put(4.,10.5){(b)}
\put(-1.5,5.5){(c)}
\put(4.,5.5){(d)}
\put(-3.8,10.8){$\etac$}
\put(-3.0,11.){$\jpsi$}
\put(-2.1,9.8){$\chiczero$}
\put(-1.4,9.3){$\chictwo$}
\end{picture}
\label{fig:ggsel}
\end{figure*}

\begin{figure*}
\includegraphics[width=0.3\textwidth]{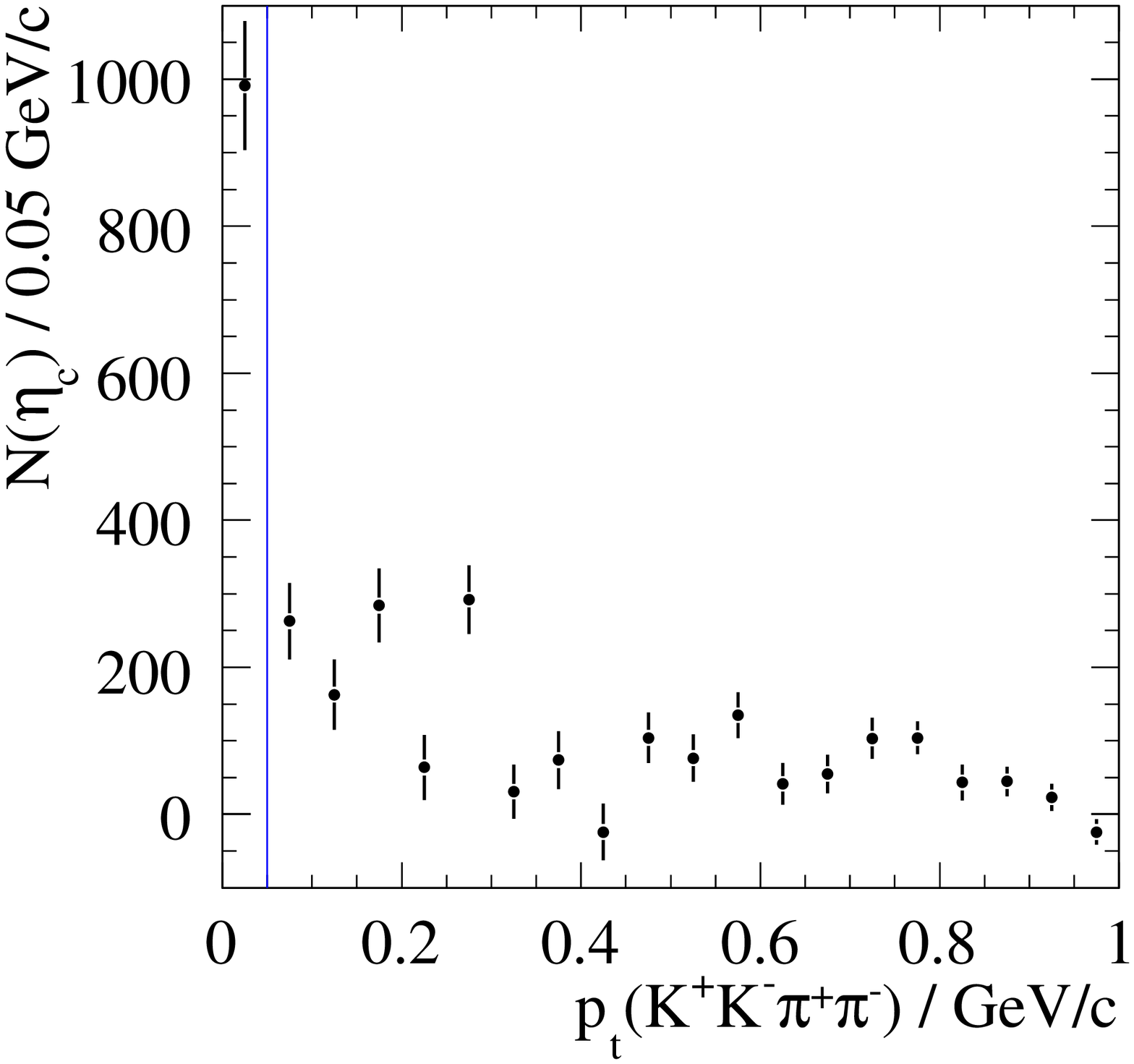}
\includegraphics[width=0.3\textwidth]{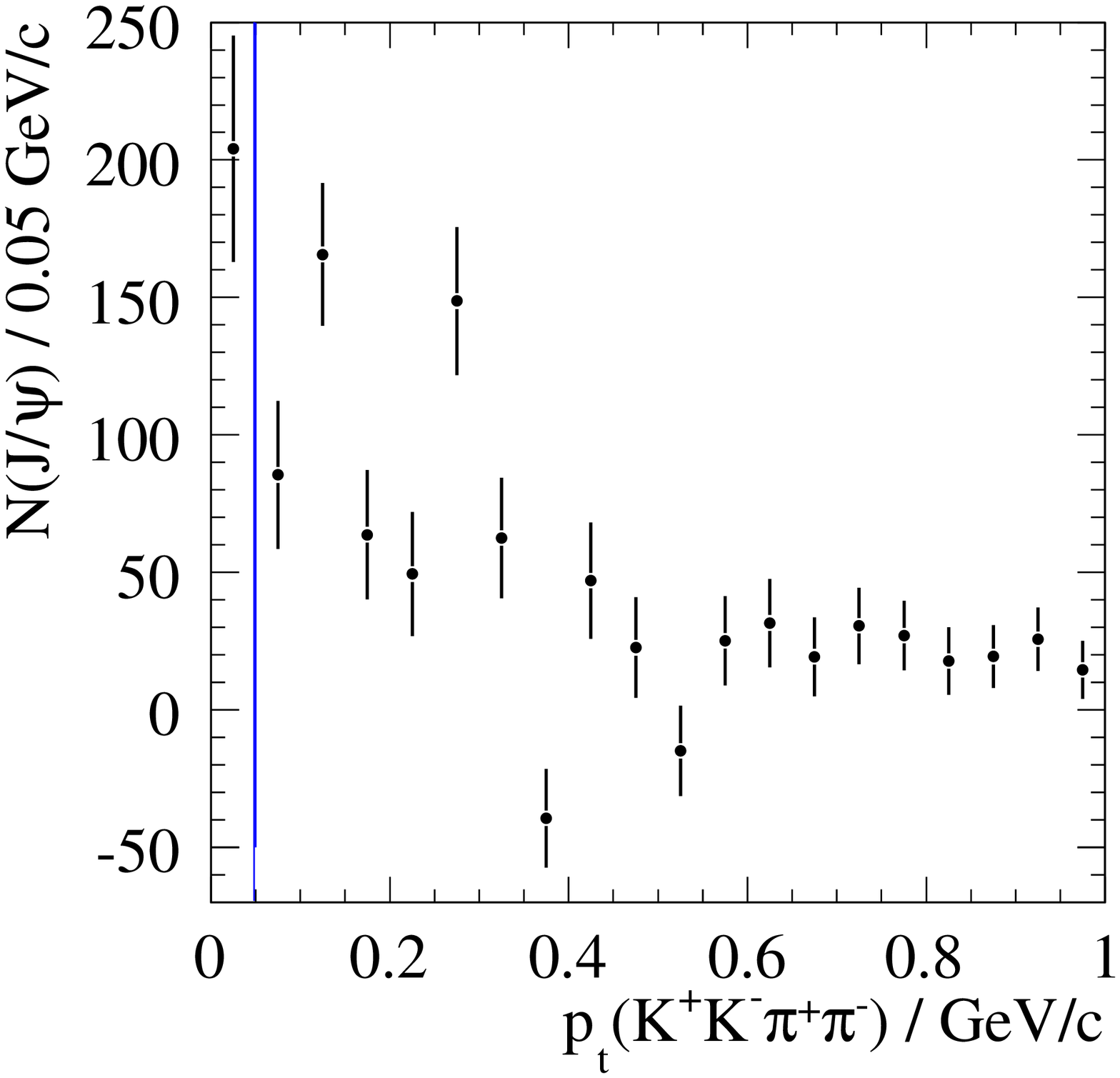}
\includegraphics[width=0.3\textwidth]{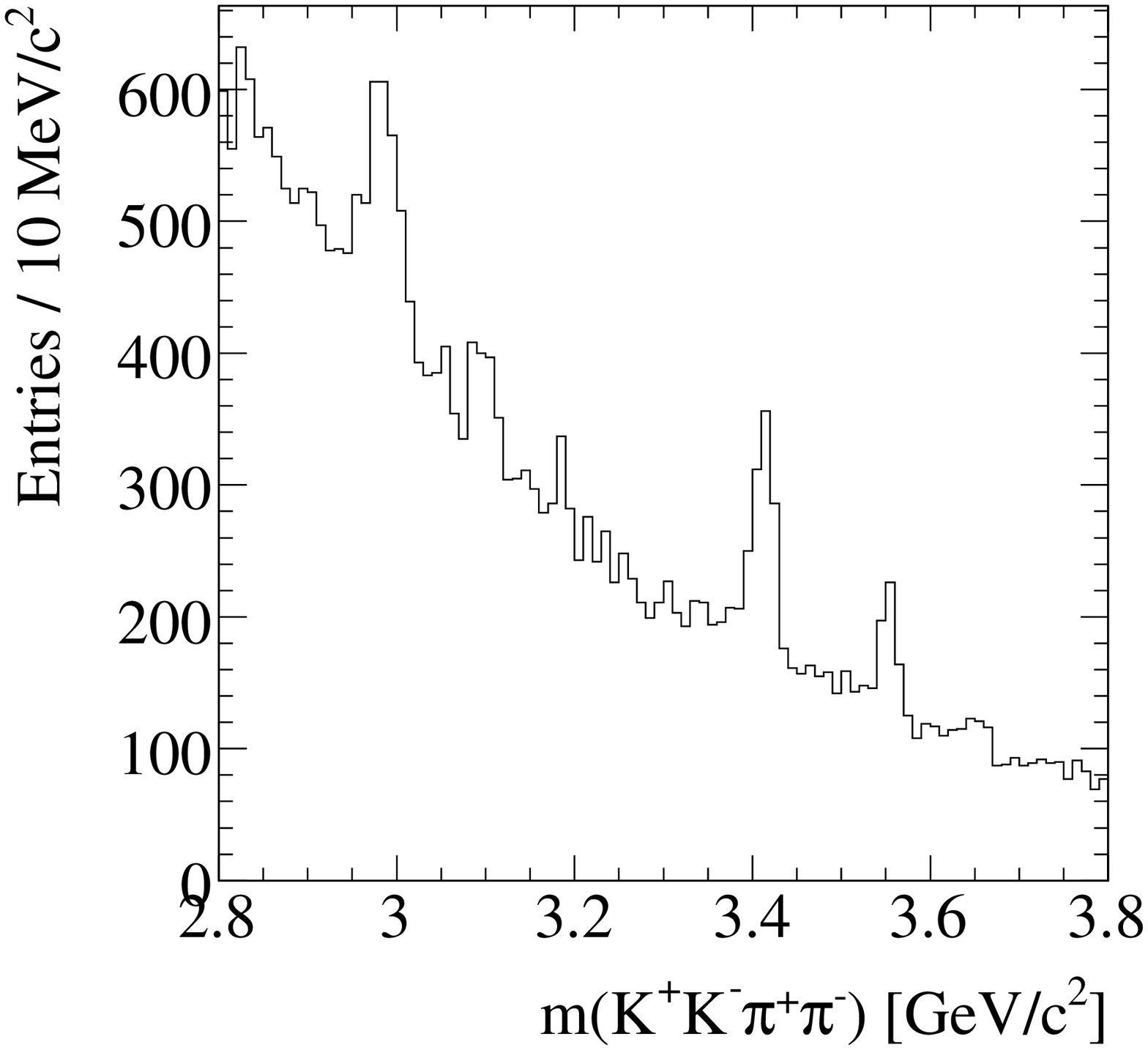}
\caption{Signal yield dependence on $p_{t}(\Kp\Km\pip\pim)$ for (a) $\etac(1S)$; (b) $\jpsi$. The vertical lines mark the $p_{t}(\Kp\Km\pip\pim) < 0.05~\mevc$ region for selecting two-photon events. (c) Resulting $\Kp\Km\pip\pim$ mass distribution after applying the principal selection criteria discussed in Sec.~\ref{sec_ggsel}.}
\begin{picture}(0,0)
\put(-4.,5.5){(a)}
\put(1.5,5.5){(b)}
\put(7.,5.5){(c)}
\put(4.9,6.3){$\etac$}
\put(5.,5.5){$\jpsi$}
\put(6.2,5.2){$\chiczero$}
\put(6.8,4.5){$\chictwo$}
\end{picture}
\label{fig:ggsel2}
\end{figure*}

The selection of two-photon-interaction events for an untagged analysis is based on established procedures (see for instance Refs.~\cite{Au04,Au09}). Due to the small scattering angles involved, most of the incoming beam energy is carried away by the $e^{+\prime}$ and $e^{-\prime}$ (see Fig.~\ref{fig:ggfeyn}). This results in a large value of the missing mass squared 
\begin{equation}
m_{\rm miss}^{2} = (p_{\ep} + p_{e^{-}} - p_{D} - p_{\Db})^{2}
\label{eq:mmass}
\end{equation}
where $p_{e^{\pm}}$ are the four-momenta of the beam electron and positron and $p_{D}$, $p_{\Db}$ are the four-momenta of the final state $D$ and $\Db$ mesons, respectively. In addition, for these events, the resultant transverse momentum of the $D\Db$ system $p_{t}(D\Db)$ is limited to small values. \\
\indent In order to establish selection criteria for $\gamma\gamma \to D\Db$ events, the reaction 
\begin{equation}
\epem \to \Kp\Km\pip\pim X
\label{eq_crea}
\end{equation}
is studied first using a data sample corresponding to an integrated luminosity of $235\invfb$. The system $X$ contains no additional charged particles. This reaction has been chosen because it has the same particle configuration as one of the final states we consider in this analysis. The charged kaons and pions are identified as described in detail in Sec.~\ref{sec_reco}. Neutral pions are reconstructed from pairs of photons with deposited energy in the EMC larger than $100\mev$. It is required that no \piz meson candidate be found in a selected event.\\
\indent Two-photon production of the $\Km\Kp\pip\pim$ system should yield large values of $m_{X}^{2}$, the missing mass squared,
\begin{equation}
m_{X}^{2} = (p_{\ep} + p_{e^{-}} - p_{\Kp} - p_{\Km} - p_{\pip} - p_{\pim})^{2}.
\label{eq:mmassX}
\end{equation}
In addition, production of the $\Km\Kp\pip\pim$ system via Initial State Radiation (ISR) should yield the small values of $m_{X}^{2}$ associated with the ISR photon, for which detection is not required. The observed distribution of the $\Km\Kp\pip\pim$ invariant mass, $m(\Km\Kp\pip\pim)$, resulting from the reaction of Eq.~(\ref{eq_crea}) is shown in Fig.~\ref{fig:ggsel}(a). \\
\indent There are clear signals corresponding to the production of $\etac(1S)$, $\chiczero(1P)$, and $\chictwo(1P)$, and, since these states all have positive $C$-parity, it is natural to associate them with two-photon production. Similarly, the large $\jpsi$ signal observed would be expected to result from ISR-production, because of the negative $C$-parity of the $\jpsi$. For the parameters of these states, see Table~\ref{tab:charmon}.\\
\indent The distribution of $m_{X}^{2}$ for $2.8 \leq m(\Km\Kp\pip\pim) \leq 3.8\gevcc$ is shown in Fig.~\ref{fig:ggsel}(b). The large peak near zero is interpreted as being due mainly to ISR production of the $\Km\Kp\pip\pim$ system, while two-photon-production events would be expected to occur at larger values of $m_{X}^{2}$. This is shown explicitly by the distributions of Figs.~\ref{fig:ggsel}(c) and~\ref{fig:ggsel}(d), which correspond to the requirements $m_{X}^{2} < 10~(\!\gevcc)^{2}$ and $m_{X}^{2} > 10~(\!\gevcc)^{2}$, respectively. \\
\indent In Fig.~\ref{fig:ggsel}(c) there is a large $\jpsi$ signal, and a much smaller $\psi(2S)$ signal can also be seen. For $\epem$ collisions at a c.m. energy $10.58\gev$, the ISR production cross section for $\jpsi$ is about three times larger than for $\psi(2S)$; also ${\cal B}(\jpsi \to \Km\Kp\pip\pim)$ is approximately nine times larger than the corresponding $\psi(2S)$ branching fraction value~\cite{Pd08}. \\
\indent It follows that the observed $\jpsi$ signal would be expected to be $\approx 27$ times larger than that for $\psi(2S)$. The signals in Fig.~\ref{fig:ggsel}(c) seem to be consistent with this expectation, and they are also in agreement with the detailed analysis of ISR production of the $\Km\Kp\pip\pim$ system in Ref.~\cite{Au07b}. There is a $\chictwo(1P)$ signal in Fig.~\ref{fig:ggsel}(c) which is comparable in size to the $\psi(2S)$ signal. The branching fraction for $\psi(2S) \to \Km\Kp\pip\pim$ is $\approx 7.5\times10^{-4}$~\cite{Pd08}, while the product ${\cal B}(\psi(2S) \to \gamma\chictwo(1P))\times {\cal B}(\chictwo(1P)\to \Km\Kp\pip\pim)$ is $\approx 7.8\times10^{-4}$~\cite{Pd08}, so that the presence of such a $\chictwo(1P)$ signal is consistent with the expected transition rates. For the $\chicone(1P)$, ${\cal B}(\psi(2S) \to \gamma\chicone(1P))\times {\cal B}(\chicone(1P)\to \Km\Kp\pip\pim) \approx 4.0\times 10^{-4}$, and so a $\chicone(1P)$ signal of approximately half the size of the $\chictwo(1P)$ signal would be expected in Fig.~\ref{fig:ggsel}(c); again the data seem to be in reasonable agreement with this expectation. \\
\indent Finally, for the $\chiczero(1P)$, ${\cal B}(\psi(2S) \to \gamma\chiczero(1P))\times {\cal B}(\chiczero(1P)\to \Km\Kp\pip\pim) \approx 16.8\times 10^{-4}$, and the corresponding signal in Fig.~\ref{fig:ggsel}(c) would be expected to be about twice the size of the $\psi(2S)$ signal. The $\chiczero(1P)$ signal seems to be larger than that of the $\psi(2S)$, but not by a factor of two; this may be because the larger energy photon from the $\psi(2S) \to \gamma\chiczero(1P)$ transition, when combined with the ISR photon, can yield a value of $m_{X}^{2}$ which is larger than $10~(\!\gevcc)^{2}$. In summary, the signals observed in Fig.~\ref{fig:ggsel}(c) appear consistent with those expected for an ISR production mechanism, especially since there is no indication of any remnant of the large $\etac(1S)$ of Fig.~\ref{fig:ggsel}(a). Furthermore, the $\chi_{cJ}$ signals in Fig.~\ref{fig:ggsel}(c) are removed by requiring that the transverse momentum of the $\Km\Kp\pip\pim$ system be less than $50\mevc$ (see discussion of Fig.~\ref{fig:ggsel}(d) below), which indicates clearly that they do not result from two-photon production.\\
\indent In Fig.~\ref{fig:ggsel}(d), the $\etac(1S)$ signal of Fig.~\ref{fig:ggsel}(a) appears to have survived the $m_{X}^{2} > 10~(\!\gevcc)^{2}$ requirement in its entirety, and the $\chiczero(1P)$ and $\chictwo(1P)$ signals have been reduced slightly, as discussed in the previous paragraph; in both Fig.~\ref{fig:ggsel}(a) and~\ref{fig:ggsel}(d) there is some indication of a small signal in the region of the $\etac(2S)$ mass. A $\jpsi$ signal of about one third of that in Fig.~\ref{fig:ggsel}(a) is present also in Fig.~\ref{fig:ggsel}(d). This is interpreted as being primarily due to a) the emission of more than one initial state photon, with the consequence that values of $m_{X}^{2}$ greater than $10~(\!\gevcc)^{2}$ are obtained, b) the ISR production of the $\psi(2S)$ with subsequent decay to $\jpsi$ + neutrals, and c) two-photon-production of the $\chictwo(1P)$ followed by $\chictwo(1P) \to \gamma \jpsi$, which has a $20\%$ branching fraction~\cite{Pd08}. \\
\indent It follows from the above that the requirement $m_{X}^{2} > 10~(\!\gevcc)^{2}$ significantly reduces ISR contributions to the $\Km\Kp\pip\pim$ final state while leaving signals associated with two-photon-production essentially unaffected. For this reason, the requirement that $m_{\mathrm{miss}}^{2}$ of Eq.~(\ref{eq:mmass}) be greater than $10~(\!\gevcc)^{2}$ is chosen as a principal selection criterion for the isolation of events corresponding to $\gamma\gamma \to D\Db$.\\
\indent As mentioned above, it is expected that for an untagged analysis of $\gamma\gamma \to D\Db$, the transverse momentum $p_{t}(D\Db)$ should be small. In order to quantify this statement, the data of Fig.~\ref{fig:ggsel}(d) were divided into intervals of $50\mevc$ in the transverse momentum of the $\Km\Kp\pip\pim$ system with respect to the $\epem$ collision axis, which is considered also to be the collision axis for two-photon-production events. For each interval a fit was made to the $m(\Km\Kp\pip\pim)$ mass distribution in the mass region $2.7 \leq m(\Km\Kp\pip\pim) \leq 3.3\gevcc$. The function used consists of a second-order polynomial to describe the background, a Gaussian function for the $\jpsi$ signal and a Breit-Wigner for the $\etac(1S)$ signal convolved with a Gaussian to account for the resolution. The $p_{t}$-dependence of the resulting $\etac(1S)$ yield is shown is Fig.~\ref{fig:ggsel2}(a), and that of the $\jpsi$ yield is shown in Fig.~\ref{fig:ggsel2}(b). The shapes of the distributions are quite similar for $p_{t} > 100\mevc$, but the interval from $50 - 100\mevc$ contains $\approx 180$ more $\etac(1S)$ signal events, and that for $0 - 50\mevc$ exhibits an excess of $\approx 800$ signal events. This behaviour is expected for two-photon-production of the $\etac(1S)$. Thus, the requirement $p_{t}(D\Db) < 50\mevc$ is imposed as the second principal selection criterion for the extraction of $\gamma\gamma \to D\Db$ events.\\
\indent Since the two-photon reactions $\gamma\gamma \to \Km\Kp\pip\pim$ and $\gamma\gamma \to D\Db$ are quasi-exclusive in the sense that only the final state $\ep$ and $e^{-}$ are undetected it is required in both instances that the total energy deposits $E_{\mathrm{EMC}}$ in the EMC which are unmatched to any charged-particle track be less than $400\mev$. The net effect is a small reduction in the smooth background. The histogram of Fig.~\ref{fig:ggsel2}(c) corresponds to the $\Km\Kp\pip\pim$ candidates of Fig.~\ref{fig:ggsel}(d) after requiring $p_{t}(\Km\Kp\pip\pim) < 50\mevc$ and that the EMC energy sum be less than $400\mev$. The $p_{t}$ criterion reduces the $\etac(1S)$ signal by a factor $\approx 2$, while the $\jpsi$ signal is reduced by a factor $\approx 5$, as is the continuum background at $2.7\gevcc$. More significantly, the continuum background at $3.7\gevcc$, just below the $D\Db$ threshold, is reduced by a factor $\approx 10$. \\
\indent It follows that the net effect of the three principal selection criteria described above (missing mass $m_{\rm miss}^{2} > 10~(\!\gevcc)^{2}$, resultant transverse momentum $p_{t}(D\Db) < 50\mevc$ and total energy deposit in the calorimeter $E_{\mathrm{EMC}}< 400\mev$) is to significantly enhance the number of two-photon-production events relative to the events resulting from ISR production, continuum production, and combinatoric background.\\
\indent Concerning the histogram of Fig.~\ref{fig:ggsel2}(c), the product $\Gamma_{\gamma\gamma}(\etac(1S))\times {\cal B}(\etac(1S)\to \Km\Kp\pip\pim)$ is $1.7 \pm 1.0$ times that for the $\chiczero(1P)$ state~\cite{Pd08}, and in Fig.~\ref{fig:ggsel2}(c) the $\etac(1S)$ signal contains $\approx 950$ events ({\it cf}.~the $0 - 50\mevc$ interval of Fig.~\ref{fig:ggsel2}(a)), while the $\chiczero(1P)$ signal contains $\approx 550$~events. It follows that the signal sizes agree well with the ratio expected on the basis of a two-photon production mechanism. In a similar vein, the ratio of the partial width $\Gamma_{\gamma\gamma}(\chi_{cJ})\times {\cal B}(\chi_{cJ} \to \Km\Kp\pip\pim)$ for $\chiczero(1P)$ and $\chictwo(1P)$ is $9 \pm 2$~\cite{Pd08}, so that after taking into account the $(2J+1)$ spin factors, the signals observed in Fig.~\ref{fig:ggsel2}(c) would be expected to be approximately in the ratio $1.8 \pm 0.4$. The $\chictwo(1P)$ signal contains $\approx 200$ events, and so is consistent with this expectation. 
\begin{table}
\caption{Charmonium states observed in the $\epem \to \Kp\Km\pip\pim X$ test data sample~\cite{Pd08}.}
\begin{ruledtabular}
\begin{tabular}{llll}
 & Mass [$\!\mevcc$] & $J^{PC}$ \\ \hline
$\etac(1S)$ & $(2980.3\pm 1.2)$ & $0^{-+}$ \\
$\jpsi(1S)$ & $(3096.916\pm 0.011)$ & $1^{--}$ \\
$\chiczero(1P)$ & $(3414.75\pm 0.31)$ & $0^{++}$ \\
$\chicone(1P)$ & $(3510.66\pm 0.07)$ & $1^{++}$ \\
$\chictwo(1P)$ & $(3556.20\pm 0.09)$ & $2^{++}$ \\
$\etac(2S)$ & $(3637\pm 4)$ & $0^{-+}$ \\
$\psi(2S)$ & $(3686.09\pm 0.04)$ & $1^{--}$ \\
\end{tabular}
\end{ruledtabular}
\label{tab:charmon}
\end{table}

\section{Reconstruction of {\boldmath $D\Db$} events}
\label{sec_reco}
\begin{figure*}
\includegraphics[width=0.3\textwidth]{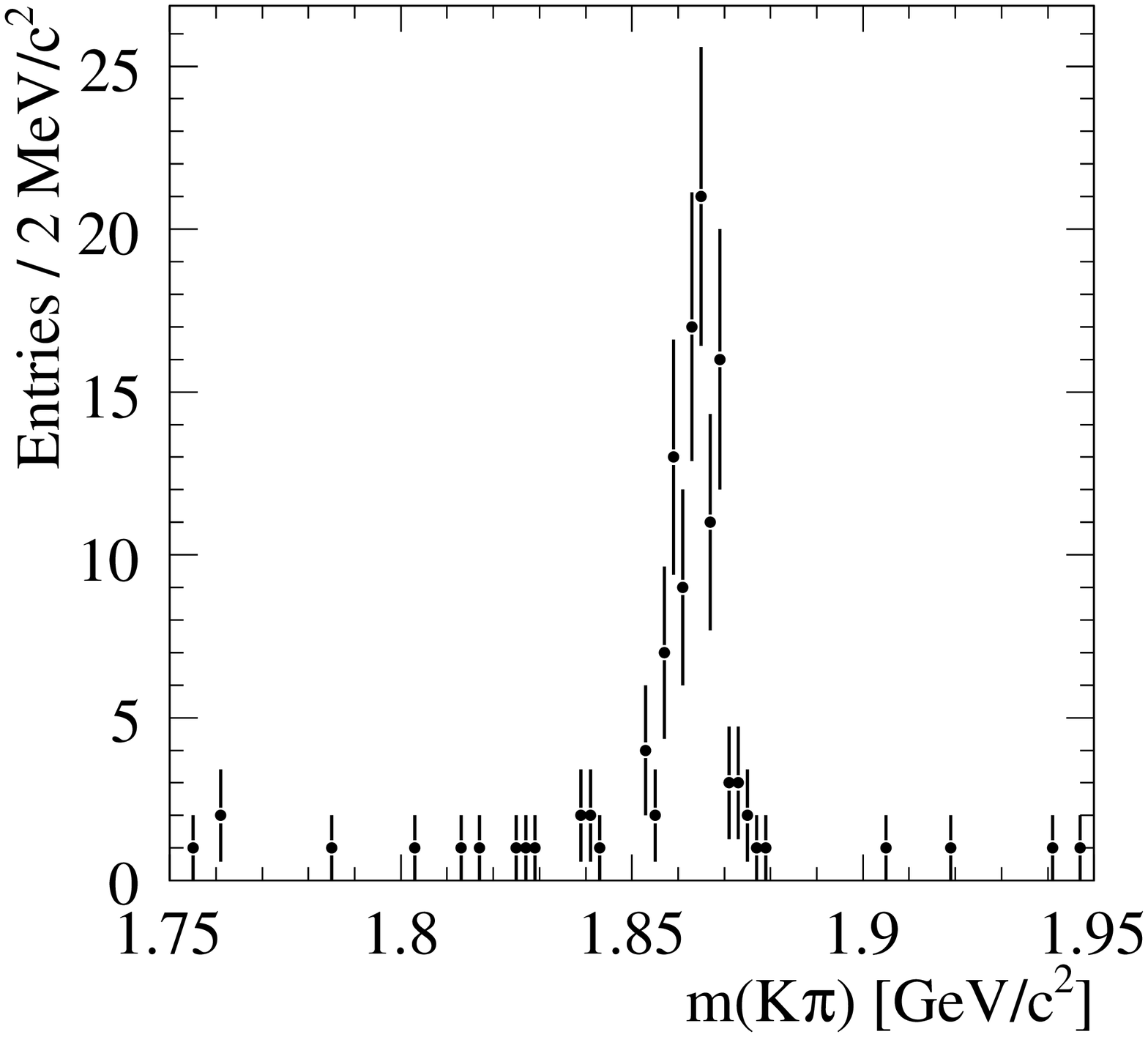}
\includegraphics[width=0.3\textwidth]{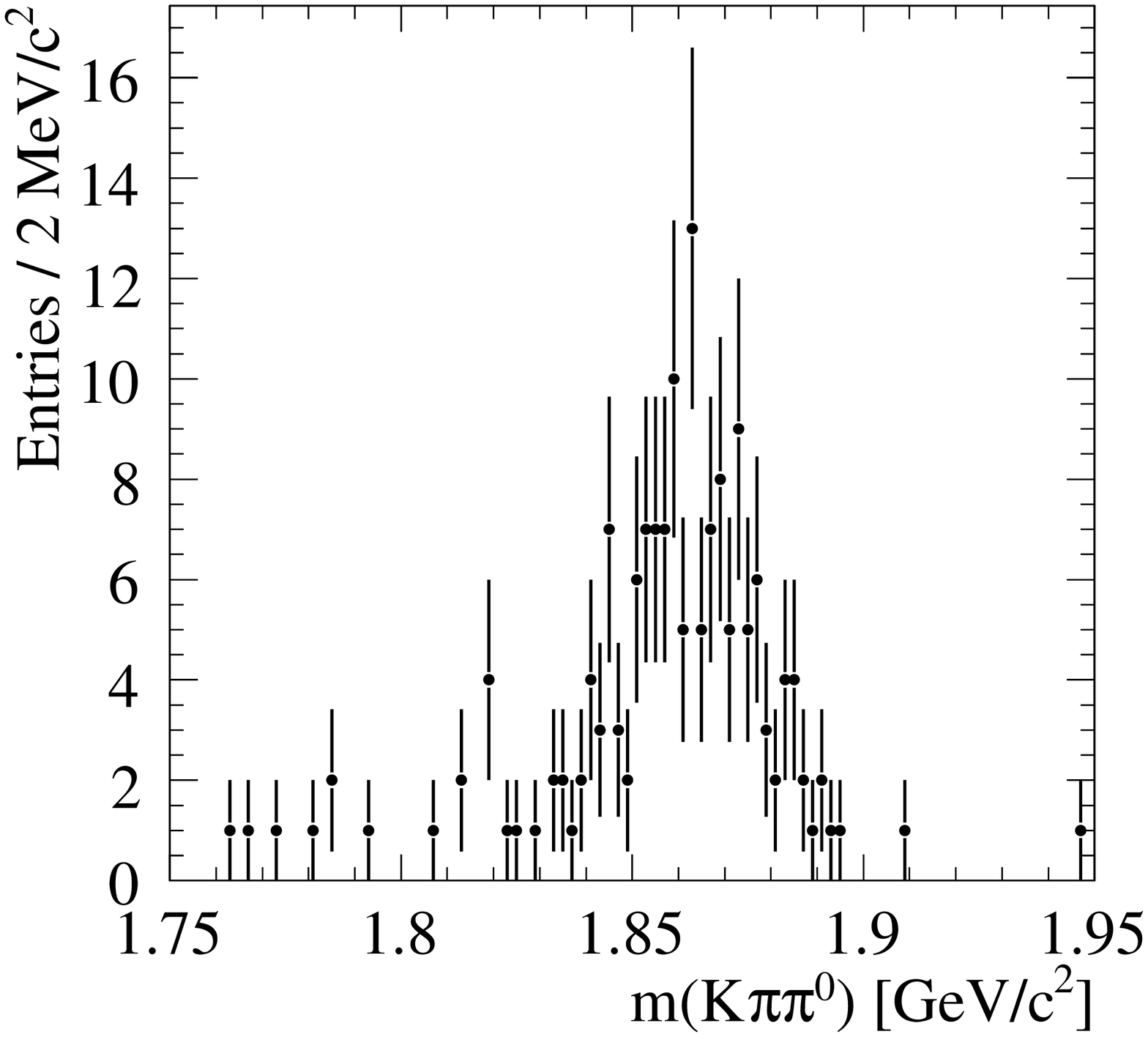}
\includegraphics[width=0.3\textwidth]{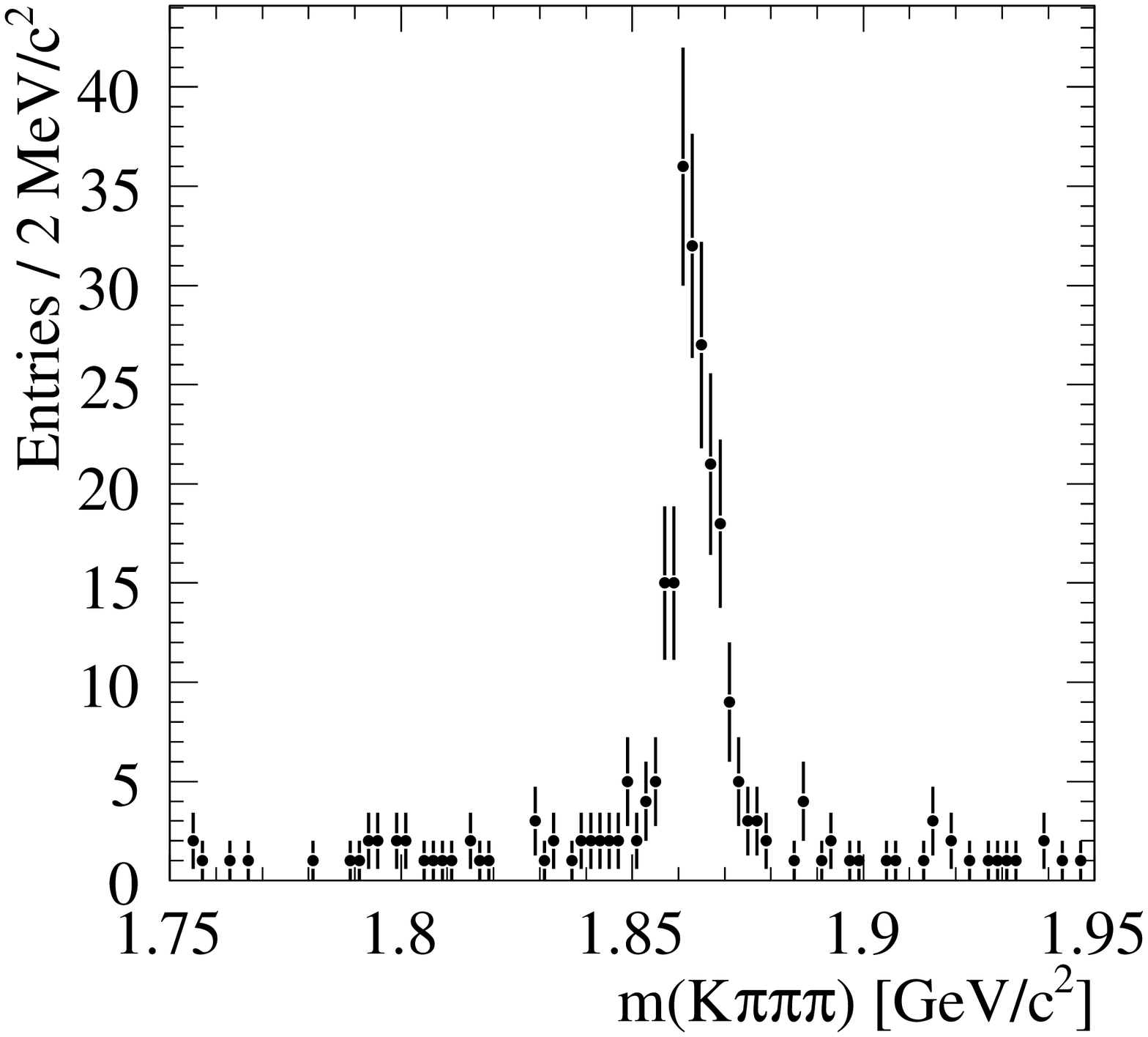}\\
\includegraphics[width=0.3\textwidth]{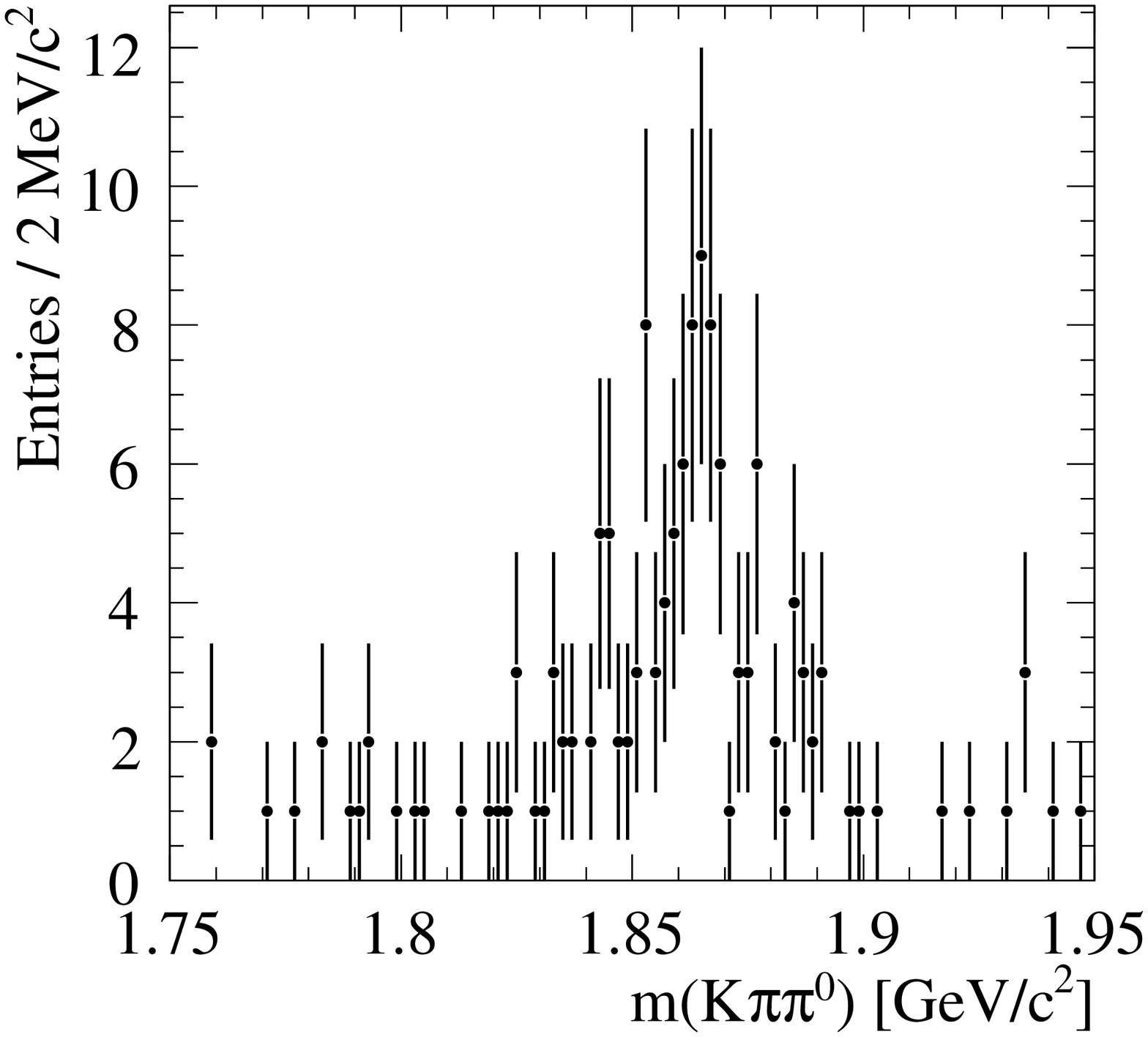}
\includegraphics[width=0.3\textwidth]{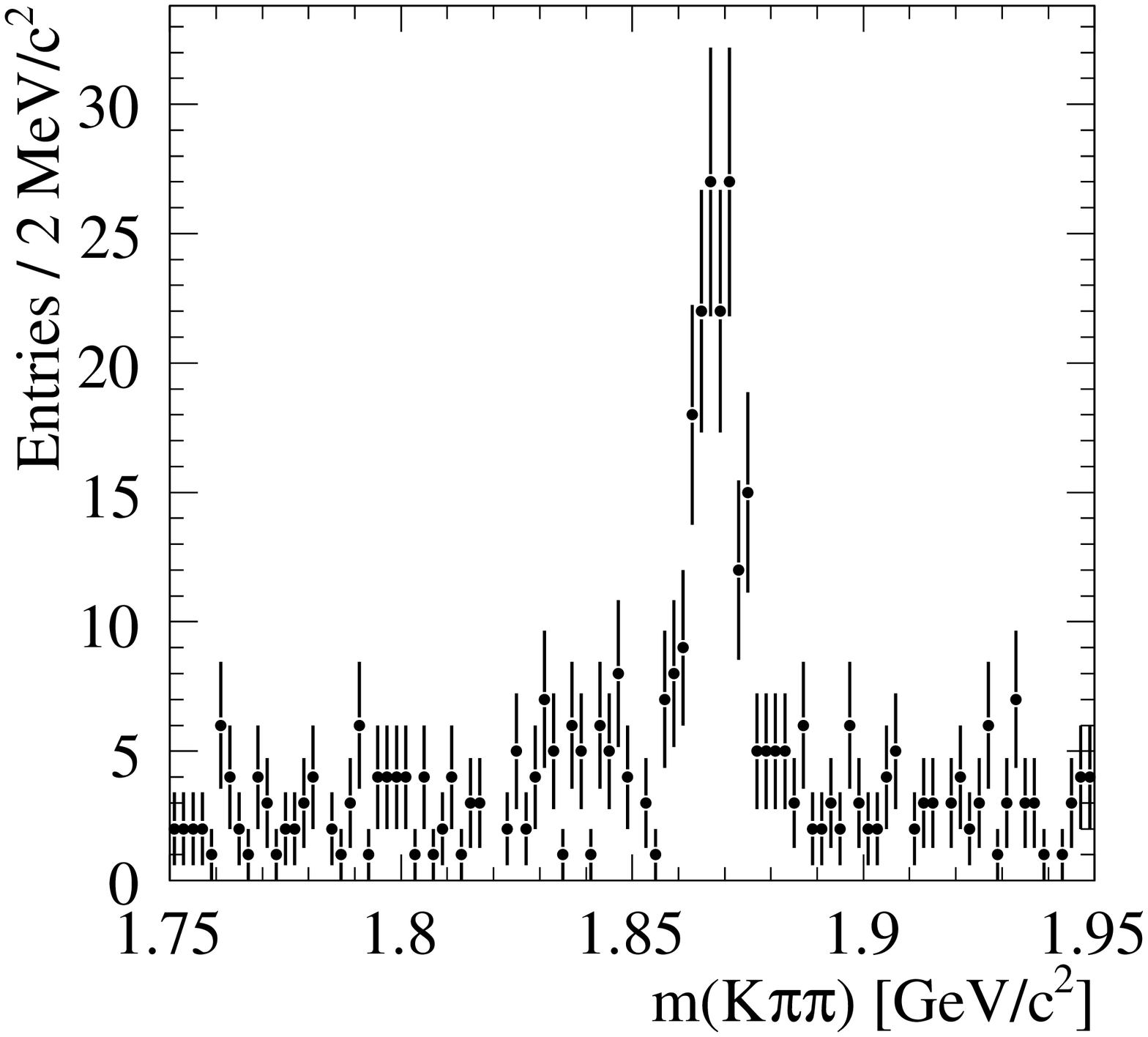}
\caption{Candidate $D$ invariant mass distributions after all selection criteria. The mass of the accompanying $\Db$ candidate is required to lie within its signal region as defined in Table~\ref{tab:sel}. (a) $\Km\pip$ in N4; (b) $\Kp\pim\piz$ in N5; (c) $\Km\pip\pip\pim$ in N6; (d) $\Kp\pim\piz$ in N7; (e) $\Km\pip\pip$ in C6.}
\label{fig:dplots}
\begin{picture}(0,0)
\put(-7.,11.){(a)}
\put(-1.5,11.){(b)}
\put(4.,11.){(c)}
\put(-4.3,5.9){(d)}
\put(1.3,5.9){(e)}
\end{picture}
\end{figure*}

\begin{figure*}
\includegraphics[width=0.3\textwidth]{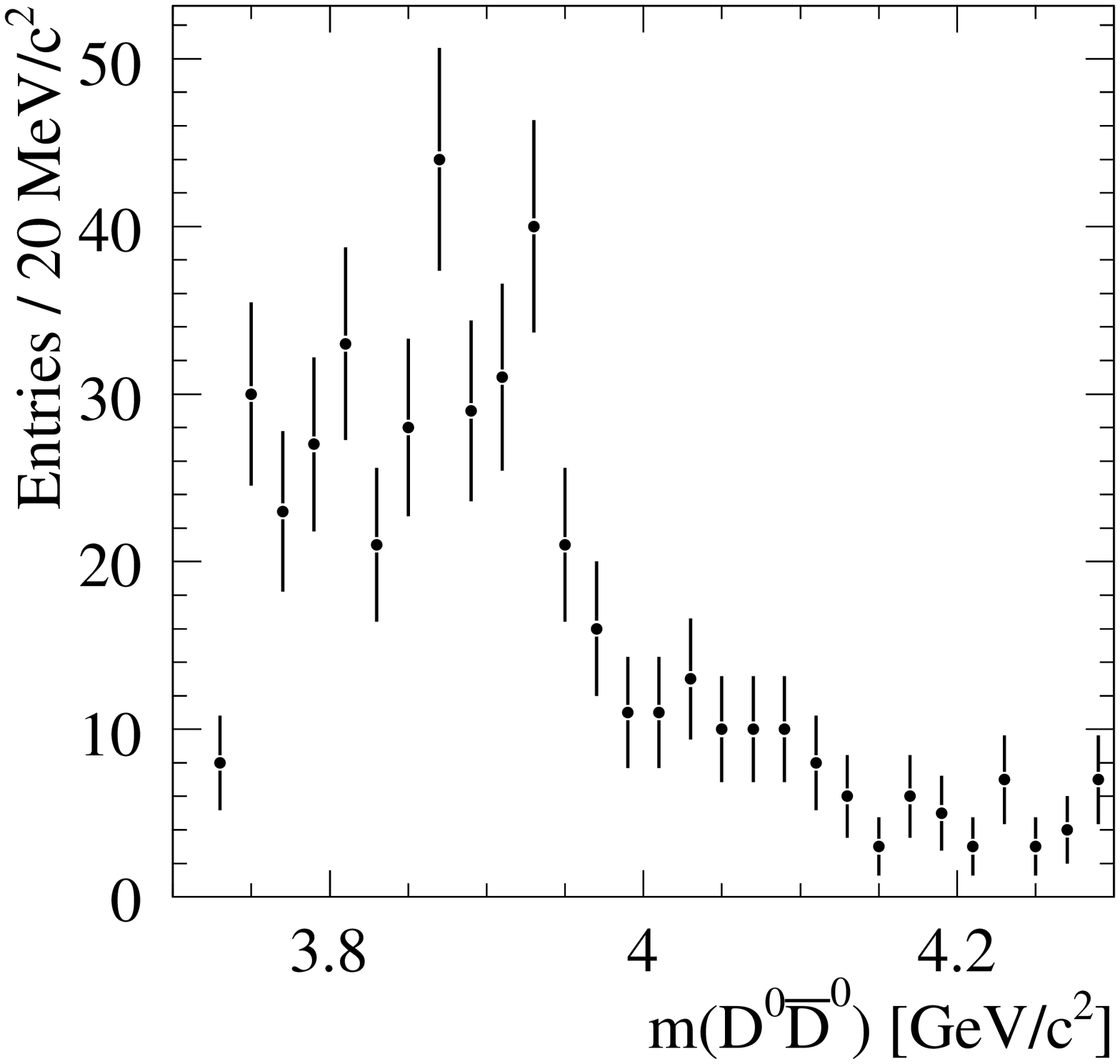}
\includegraphics[width=0.3\textwidth]{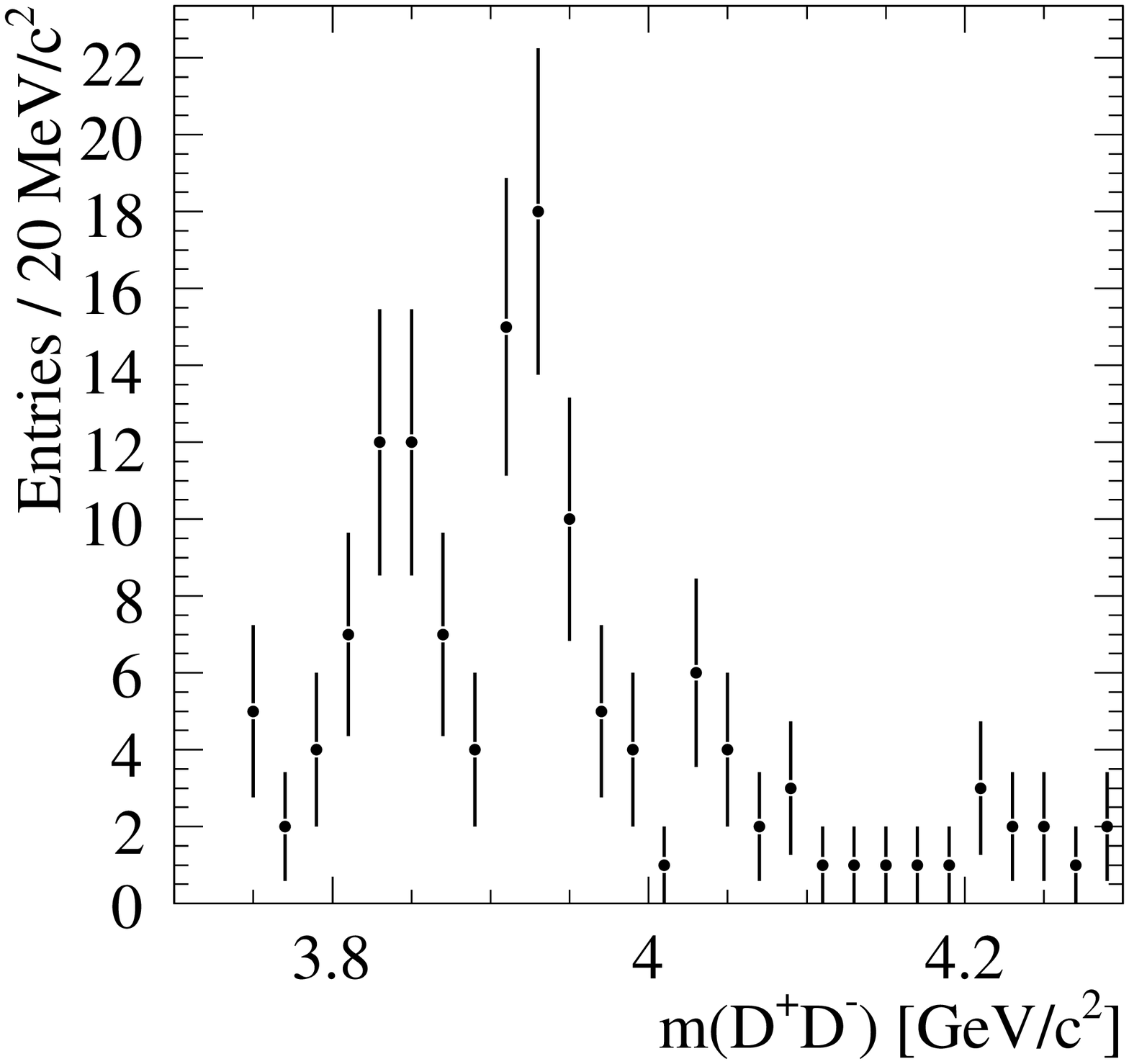}
\includegraphics[width=0.3\textwidth]{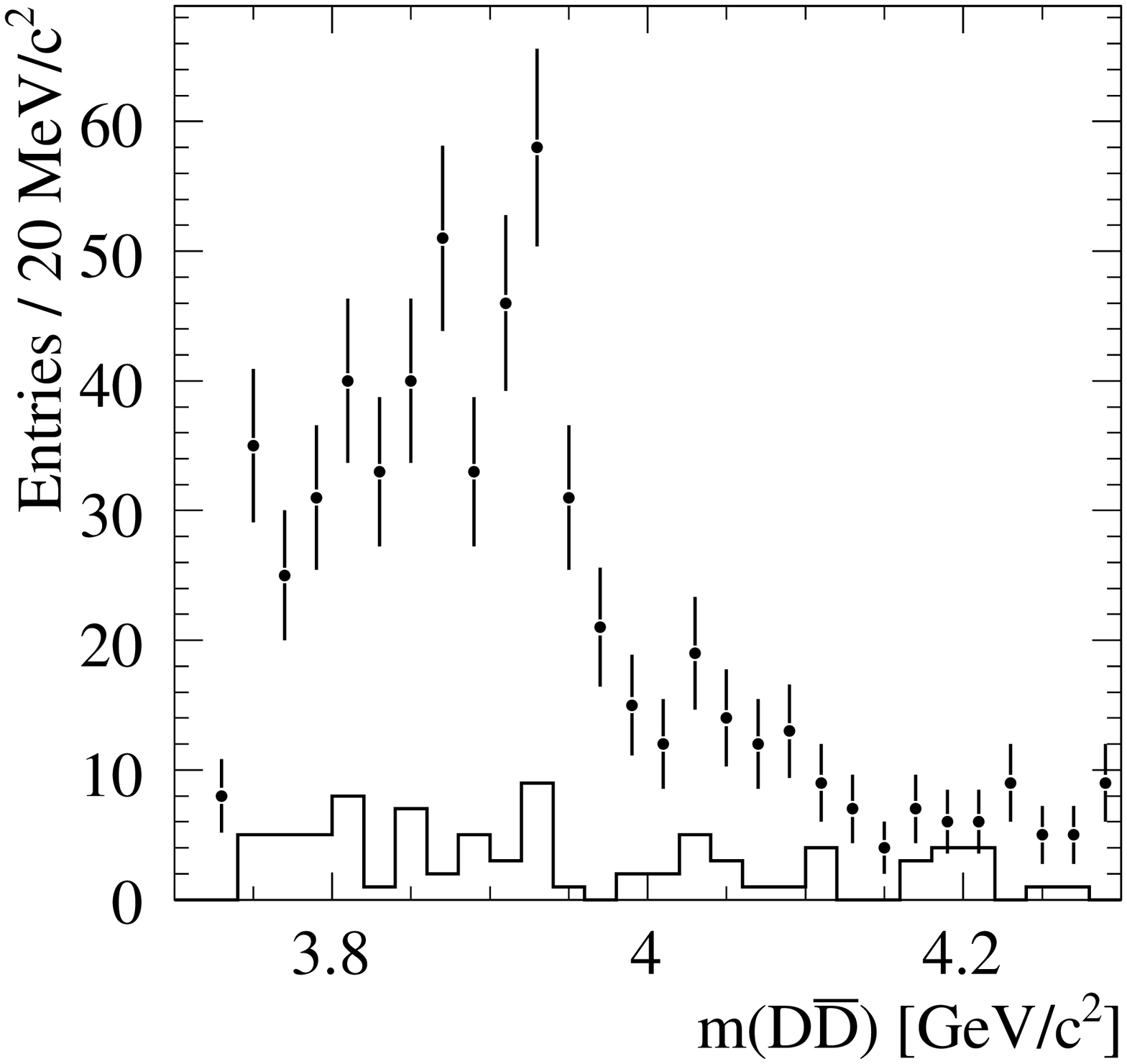}
\caption{$D\Db$ invariant mass distribution for the (a) $\DzDzb$ and (b) $\Dp\Dm$ channels. (c) The combined $D\Db$ invariant mass distribution. The open histogram in (c) shows the combinatoric background estimated from the $D$-mass sidebands.}
\label{fig:ddmass}
\begin{picture}(0,0)
\put(-4.,5.5){(a)}
\put(1.5,5.5){(b)}
\put(7.,5.5){(c)}
\end{picture}
\end{figure*}

\begin{figure}
\includegraphics[width=0.3\textwidth]{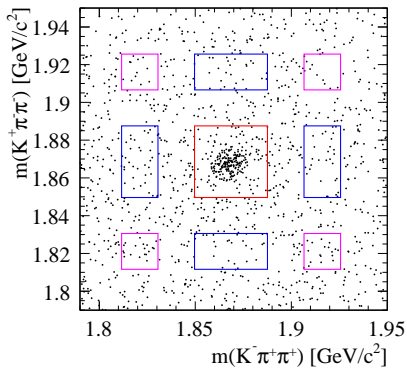}
\caption{$\Kp\pim\pim$ mass {\it vs} $\Km\pip\pip$ mass for the channel C6 of Table~\ref{tab:dmodes}. The boxes correspond to the signal and sideband regions.}
\label{fig:ntile}
\end{figure}

Candidate $D\Db$ events are reconstructed in the five combinations of $D$ decay modes listed in Table~\ref{tab:dmodes} (the use of charge conjugate states is implied throughout the text). Events are selected by requiring the exact number of charged-particle tracks defined by the relevant final state. \\
\indent Track selection requirements include transverse momentum $p_{t} > 0.1\gevc$, at least 12 coordinate measurements in the DCH, a maximum distance of closest approach (DOCA) of $1.5 \cm$ to the $z$-axis, with this point at a maximum DOCA of $10 \cm$ to the $xy$-plane at $z=0$. \\
\indent Kaon candidates are identified based on the normalized kaon, pion and proton likelihood values ($L_{K}$, $L_{\pi}$ and $L_{p}$) obtained from the particle identification system, by requiring $L_{K}/(L_{K}+L_{\pi}) > 0.9$ and $L_{K}/(L_{K}+L_{p}) > 0.2$. Tracks that fulfill $L_{K}/(L_{K}+L_{\pi}) < 0.82$ and $L_{p}/(L_{p}+L_{\pi}) < 0.98$ are selected as pions. Additionally, in both cases the track should be inconsistent with electron identification.\\
\begin{table}
\caption{$D$ decay final states studied in this analysis; for channels N5, N6, and N7, inclusion of the corresponding charge conjugate combination is implied.}
\begin{ruledtabular}
\begin{tabular}{llll}
\multicolumn{2}{l}{Channel} & $D$ decay mode & $\Db$ decay mode \\ \hline
N4 & $\DzDzb$ & $\Dz \to \Km\pip$ & $\Dzb \to \Kp\pim$ \\
N5 & $\DzDzb$ & $\Dz \to \Km\pip$ & $\Dzb \to \Kp\pim\piz$ \\
N6 & $\DzDzb$ & $\Dz \to \Km\pip$ & $\Dzb \to \Kp\pim\pim\pip$ \\
N7 & $\DzDzb$ & $\Dz \to \Km\pip\pip\pim$ & $\Dzb \to \Kp\pim\piz$ \\
C6 & $\Dp\Dm$ & $\Dp \to \Km\pip\pip$ & $\Dm \to \Kp\pim\pim$ \\
\end{tabular}
\end{ruledtabular}
\label{tab:dmodes}
\end{table}
\indent Photon candidates are selected when their deposited energy in the EMC is larger than $100\mev$. Neutral pions are reconstructed from pairs of photons with combined mass within $[0.115,0.155]~\gevcc$ and a $\piz$ mass constraint is applied to them. \\
\indent The $D$ candidate decay products are fitted to a common vertex with a $D$ meson mass constraint applied; candidates with a $\chi^{2}$ fit probability greater than $0.1 \%$ are retained. Accepted $D\Db$ pairs are refitted to a common vertex consistent with the $\epem$ interaction region, and those with a $\chi^{2}$ fit probability $p_{v}(D\Db)$ greater than $0.1 \%$ are retained. Events with $\piz$ candidates other than those from a $D$ or $\Db$ decay of interest are rejected. These preselection criteria are identical for all five combinations of $D$ decay modes.\\
\indent The signal regions for accepted, unconstrained $D$ candidates are then fitted using a multi-Gaussian signal function 
\begin{equation}
 R(m)=\int_{\sigma_0}^{r\sigma_0} \frac{1}{r\sigma^2}e^{-\frac{(m-m_{0})^2}{2\sigma^2}}\mathrm{d}\sigma
\label{eq:mulg}
\end{equation}
with free parameters $\sigma_{0}$, $r$ (minimal and maximal width) and $m_{0}$; the background is described by a polynomial. The full width at half maximum ($\mathrm{FWHM}$) of the signal lineshape in data is used to define each $D$ signal region; $D$ candidates are selected from a region of width $\pm 1.5~\mathrm{FWHM}$ around the mean mass. The mass windows are listed in Table~\ref{tab:sel}. \\
\indent From the list of accepted $D\Db$ candidates those produced in two-photon events are then selected by applying the three criteria defined in Sect.~\ref{sec_ggsel} (summarized in Table~\ref{tab:sel2}). These criteria are also identical for all combinations of $D$ decay modes.  \\
\indent Depending on the decay mode, up to $2.5\%$ of the events have multiple candidates which passed all selection criteria. In this case, the candidate with the best fit probability $p_{v}(D\Db)$ is chosen. Based on MC studies, the correct candidate is selected in more than $99 \%$ of the cases with this method. \\
\indent The resulting invariant mass spectra for $D$ meson candidates after all selection criteria have been applied are shown in Fig.~\ref{fig:dplots} for events in which the mass of the recoil $\Db$ candidate lies within the defined signal region. In all modes, clear signals with small backgrounds are obtained. The resulting $D\Db$ invariant mass distributions are shown in Fig.~\ref{fig:ddmass}(a) and~\ref{fig:ddmass}(b) for the neutral modes (N4, N5, N6, N7), and for the charged mode (C6), respectively. The combined spectrum is shown in Fig.~\ref{fig:ddmass}(c). An enhancement near $3.93\gevcc$ is visible.\\
\indent To estimate the amount of combinatoric background in the signal region, the two-dimensional space spanned by the invariant masses of the $D$ and $\Db$ candidates is divided into nine regions: one central signal region and eight sideband regions above and below the signal region as shown in Fig.~\ref{fig:ntile} for the $\Km\pip\pip/\Kp\pim\pim$ (C6) mode. The mass range for the signal region is $\pm~1.5~\mathrm{FWHM}$ around the mean mass. The sideband regions are $1.5~\mathrm{FWHM}$ wide, leaving a gap of $1.5~\mathrm{FWHM}$ between signal and sideband. No significant contribution from combinatoric background is observed in the $D\Db$ spectrum (Fig.~\ref{fig:ddmass}(c)).\\
\indent An attempt was made to isolate the signal in Fig.~\ref{fig:ddmass}(c) by a weighting method. This assumes that signal and background events have different angular distributions, and was successfully used in a previous \babar~analysis~\cite{DuZ07}. Simulations with a $J^{PC}=2^{++}$ signal (generated with its correct angular distribution) plus background showed that the method works well with high signal statistics and moderate background, but is not reliable with the limited statistics and background of the current analysis. Therefore, the method was not considered further in the present analysis.

\begin{table}
\caption{Summary of selection criteria used for identifying $D\Dbar$ candidates.}
\begin{ruledtabular}
\begin{tabular}{llll}
Channel & number & $D$ mass window & $\Db$ mass window \\ 
& of \piz & [$\!\mevcc$] & [$\!\mevcc$] \\ \hline
N4 & 0 & $1863.4 \pm 22$ & $1863.4 \pm 22$  \\ 
N5 & 1 & $1863.4 \pm 22$ & $1863.4 \pm 43$  \\  
N6 & 0 & $1863.4 \pm 22$ & $1863.4 \pm 16$  \\ 
N7 & 1 & $1863.4 \pm 16$ & $1863.4 \pm 43$  \\ 
C6 & 0 & $1868.5 \pm 18$ & $1868.5 \pm 18$  \\  
\end{tabular}
\end{ruledtabular}
\label{tab:sel}
\end{table}

\begin{table}
\caption{Summary of requirements used for selecting only $D\Dbar$ candidates from two-photon events. These criteria are identical for all decay modes.}
\begin{ruledtabular}
\begin{tabular}{llll}
Channel & $m_{\mathrm{miss}}^2$ & $p_{t}(D\Db)$ & $E_{\mathrm{EMC}}$\\ 
& [$(\!\gevcc)^{2}$] & [$\!\gevc$] & [$\!\gev$] \\ \hline
all modes & $> 10.0$ & $<0.05$ & $<0.4$  \\ 
\end{tabular}
\end{ruledtabular}
\label{tab:sel2}
\end{table}

\begin{figure}
\includegraphics[width=0.3\textwidth]{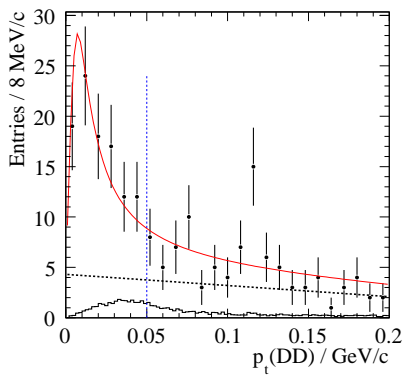}
\caption{Distribution of $p_{t}(D\Db)$ for data in the $Z(3930)$ signal region. The fitted lineshape consists of the expected $\gamma\gamma$ lineshape obtained from MC plus a linear background (dotted line). The vertical line shows the $p_{t}$ criterion for selecting $\gamma\gamma$ events. The histogram shows the shape of the $p_{t}(D\Db)$ distribution from simulated $\Dstar\Db$ events with missing $\piz$ or $\gamma$. The bump in this distribution is not seen in the data distribution, indicating that any $\Dstar\Db$ background is small.}
\label{fig:pt}
\end{figure}

\begin{figure}
\includegraphics[width=0.3\textwidth]{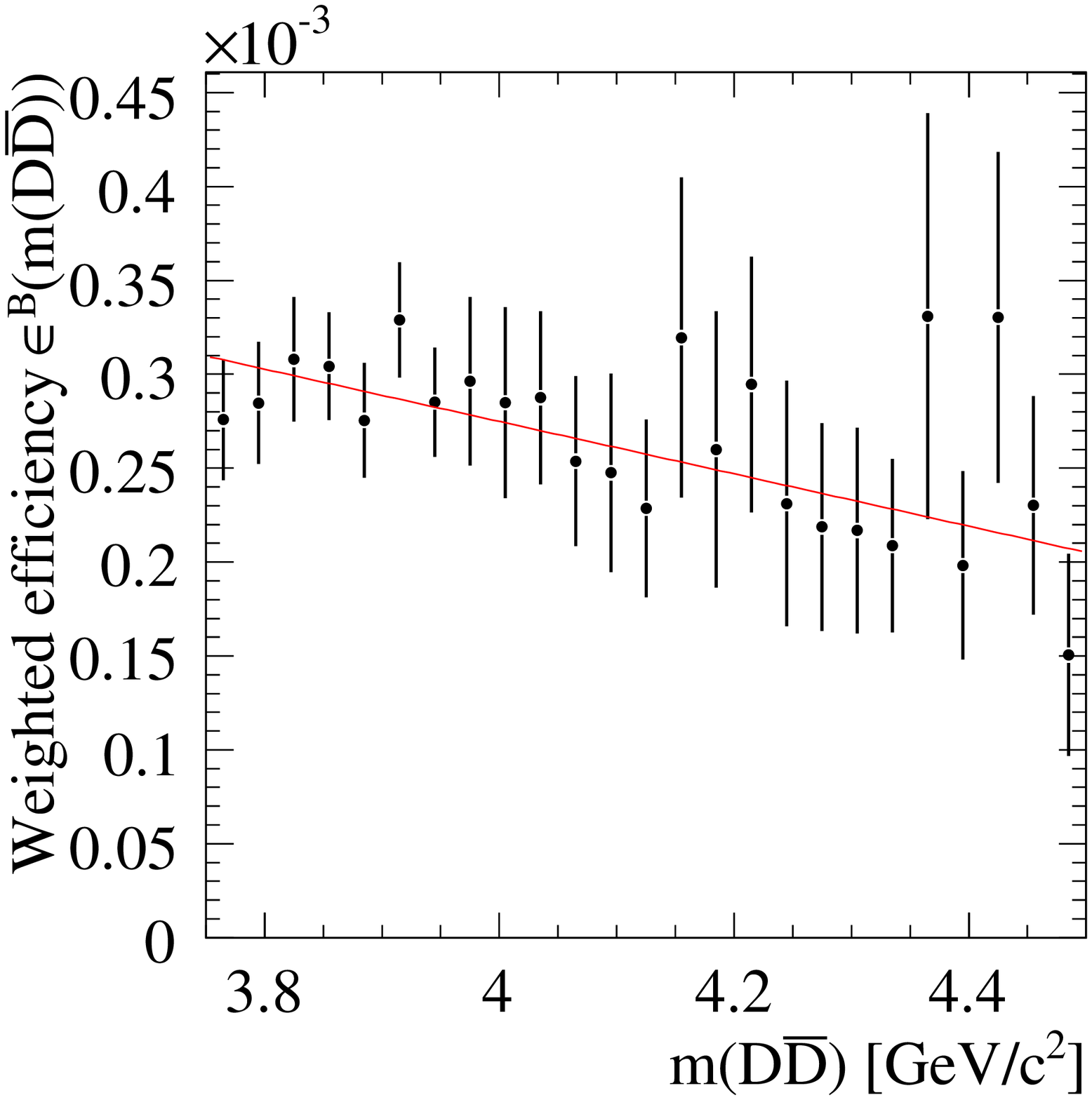}
\caption{Mass dependence of the weighted reconstruction efficiency calculated using Eq.~(\ref{eq:meaneff}), described by a straight line.}
\label{fig:effi}
\end{figure}

\section{Monte Carlo Studies}
\label{sec_mcstud}

For modeling the detector resolution, efficiency studies and the estimation of the two-photon width $\Gamma_{\gamma\gamma}$ of the resonance, Monte Carlo (MC) events were generated which pass the same reconstruction and analysis chain as the experimental data. For each signal decay channel about $10^{6}$ events were generated. Additional events were generated for background modes involving $\Dstar$ mesons. The {\tt GamGam} two-photon event generator was used to simulate $\gamma\gamma \to Z(3930) \to D\Db$ events, while the decays of the $D$ and $\Db$ mesons were generated by {\tt EvtGen}~\cite{La01}. The detector response was simulated using the {\tt GEANT4}~\cite{Ag03} package. The program {\tt GamGam} uses the BGMS formalism~\cite{Bu75}. It was developed for CLEO and was used for example in the analysis of $\chiczero(1P),\chictwo(1P) \to 4\pi$ decays~\cite{Ei01}. {\tt GamGam} was later adapted to \babar~and used for the analysis of $\etac(1S,2S) \to K_{S}^{0}K^{\pm}\pi^{\mp}$~\cite{Au04}.\\
\indent 
For small photon virtualities $|q_{1,2}|^{2}$ (see Fig.~\ref{fig:ggfeyn}) the differential cross section for the process $\epem \to \epem\gamma\gamma, \gamma\gamma \to X$ is given by the product $L\times F \times \sigma(\gamma\gamma\to X)$, where $L$ is the two-photon flux. The form factor $F$ extrapolates the process to virtual photons and is {\it a priori} not known. A plausible model
\begin{equation}
  F = \left(\frac{1}{1-q_{1}^{2}/m_{v}^{2}}\right)^{2}\times\left(\frac{1}{1-q_{2}^{2}/m_{v}^{2}}\right)^{2}
\label{eq_ff}
\end{equation}
is used~\cite{Po86}, with $m_{v}$ being the mass of an appropriate vector boson ($\rho$, $\jpsi$, $Z^{0}$). In the calculations relevant to this analysis $m_{v} = m(\jpsi)$ was used, as the $Z(3930)$ is expected to be a charmonium state. An alternative model was used in order to evaluate systematic uncertainties associated with MC simulations (see Section~\ref{sec_syst}).\\
\indent To validate the {\tt GamGam} generator its output was compared to that of another two-photon generator ({\tt TREPS}) used by Belle~\cite{Ue96}. The cross sections for the reactions  $\epem\to \epem\gamma\gamma, \gamma\gamma \to \etac(1S,2S)$  were calculated in {\tt GamGam} and compared to the Belle values~\cite{Ue08}. In order to compare the different generators, the cross sections were calculated using the hypothetical values $\Gamma_{\gamma\gamma}\times {\cal B}(\etac(1S,2S)\to {\rm final~state}) = 1\kev$, and $q_{1,2}^{2}$ was restricted to values smaller than $1~(\!\gevcc)^{2}$. The {\tt TREPS} results were $2.11~\mathrm{pb}$ for $\etac(1S)$ and $0.86~\mathrm{pb}$ for $\etac(2S)$. The corresponding {\tt GamGam} values were $2.13~\mathrm{pb}$ and $0.84~\mathrm{pb}$, respectively. The two generators are in agreement at the level of a few percent.\\
\indent For a global check, the cross sections for the continuum reaction $\epem\to \epem\gamma\gamma, \gamma\gamma \to \mup\mun$ were calculated with {\tt GamGam} for various CM energies and compared to QED predictions~\cite{Be84,Ber84}, which describe the data with high accuracy~\cite{Ad82}. Here, the agreement was slightly worse, due to the imperfect tuning of the {\tt GamGam} program for these reactions. Similar results were obtained when checking against calculations with non-relativistic models for $\etac(1S)$\ and $\chictwo(1P)$~\cite{Sc98}. Nevertheless this comparison showed that {\tt GamGam} works properly under these conditions also. These studies lead to the assignment of a total systematic uncertainty of $\pm 3~\%$ associated with the MC simulation (see Sec.~\ref{sec_syst}).

\section{\boldmath{Purity of the $\gamma\gamma \to D\Db$ sample}}
\label{sec_puri}
The selection criteria used to enhance the two-photon content of the $D\Db$ sample were discussed in Sec.~\ref{sec_ggsel}. They were developed by investigating the reaction $\epem \to \Kp\Km\pip\pim X$ of Eq.~(\ref{eq_crea}). Figure~\ref{fig:ggsel2}(c) shows that after the selection procedure the signals associated with $\gamma\gamma$-reactions, like that for the $\etac(1S)$, are enhanced, while signals such as that for the $\jpsi$, which are typical of ISR production, are suppressed. The $p_{t}(D\Db)$ distribution is shown in Fig.~\ref{fig:pt} for events in the $Z(3930)$ signal region, defined as the region from $3.91$ to $3.95\gevcc$. Here the $p_{t}(D\Db)$ selection criterion has not been applied. The data are fitted with a curve for $\gamma\gamma$ events obtained from MC, plus a linear background derived from sideband studies of the $D\Db$ mass spectrum. The fit indicates that the majority of $D\Db$ candidates in the signal region result from two-photon interactions. 

\section{Reconstruction efficiency}
\label{sec_effi}
The reconstruction efficiency for each decay mode is calculated as a function of $m(D\Db)$ using MC events which pass the same reconstruction and selection criteria as real events and includes detector acceptance, track reconstruction- and particle identification efficiencies. The mass-dependent efficiency $\epsilon_{i}(m(D\Db))$ for each channel $i$ is fitted with a polynomial in $m(D\Db)$ and is found in each case to decrease with increasing $D\Db$ mass. For the combination of modes (Fig.~\ref{fig:ddmass}(c)), an overall weighted efficiency $\epsilon^{B}(m(D\Db))$, which includes the branching fractions for the $D$ decays, is computed using
\begin{equation}
\epsilon^{B}(m(D\Db)) = \frac{5}{2}\frac{\sum_{i=1}^{5}N_{i}(m(D\Db))}{\sum_{i=1}^{5}\frac{N_{i}(m(D\Db))}{\epsilon_{i}^{B}(m(D\Db))}},
\label{eq:meaneff}
\end{equation}
as was done in Ref.~\cite{Au07}; $N_{i}(m(D\Db))$ is the number of $D\Db$ candidates in the data mass spectrum for channel $i$, and $\epsilon_{i}^{B}(m(D\Db))$ is defined as the product of the efficiency $\epsilon_{i}$ as parameterized by the fitted polynomial and the branching fraction ${\cal B}_{i}$ for the $i$-th channel, as follows
\begin{equation}
\epsilon_{i}^{B}(m(D\Db)) = \epsilon_{i}(m(D\Db))\times {\cal B}_{i}.
\label{eq:eff}
\end{equation}
The factor $\frac{1}{2}$ originates from referring to $D\Db$ ($\DzDzb$ and $\Dp\Dm$) events; the factor 5 from summing over the five channels. Figure~\ref{fig:effi} shows the mass dependence of $\epsilon^{B}(m(D\Db))$, which is parameterized by a straight line. The large uncertainties are due to the limited statistics available in the data samples. The error bars do not contain the uncertainties in the branching fractions; these will be discussed separately in Sec.~\ref{sec_syst} in the context of systematic error estimation. The data are weighted by this mean efficiency, which is scaled by a constant value $d$ to obtain weights near one,
\begin{equation}
\epsilon(m(D\Db)) = d\times \epsilon^{B}(m(D\Db))
\label{eq:effscale}
\end{equation}
as weights far from one might result in incorrect errors for the signal yield obtained in the maximum likelihood fit~\cite{Fr79}. The resulting $D\Db$ mass distribution will be discussed in Sec.~\ref{sec_fit}.

\section{Detector resolution and signal yield}
\label{sec_fit}

\begin{figure*}
\includegraphics[width=0.3\textwidth]{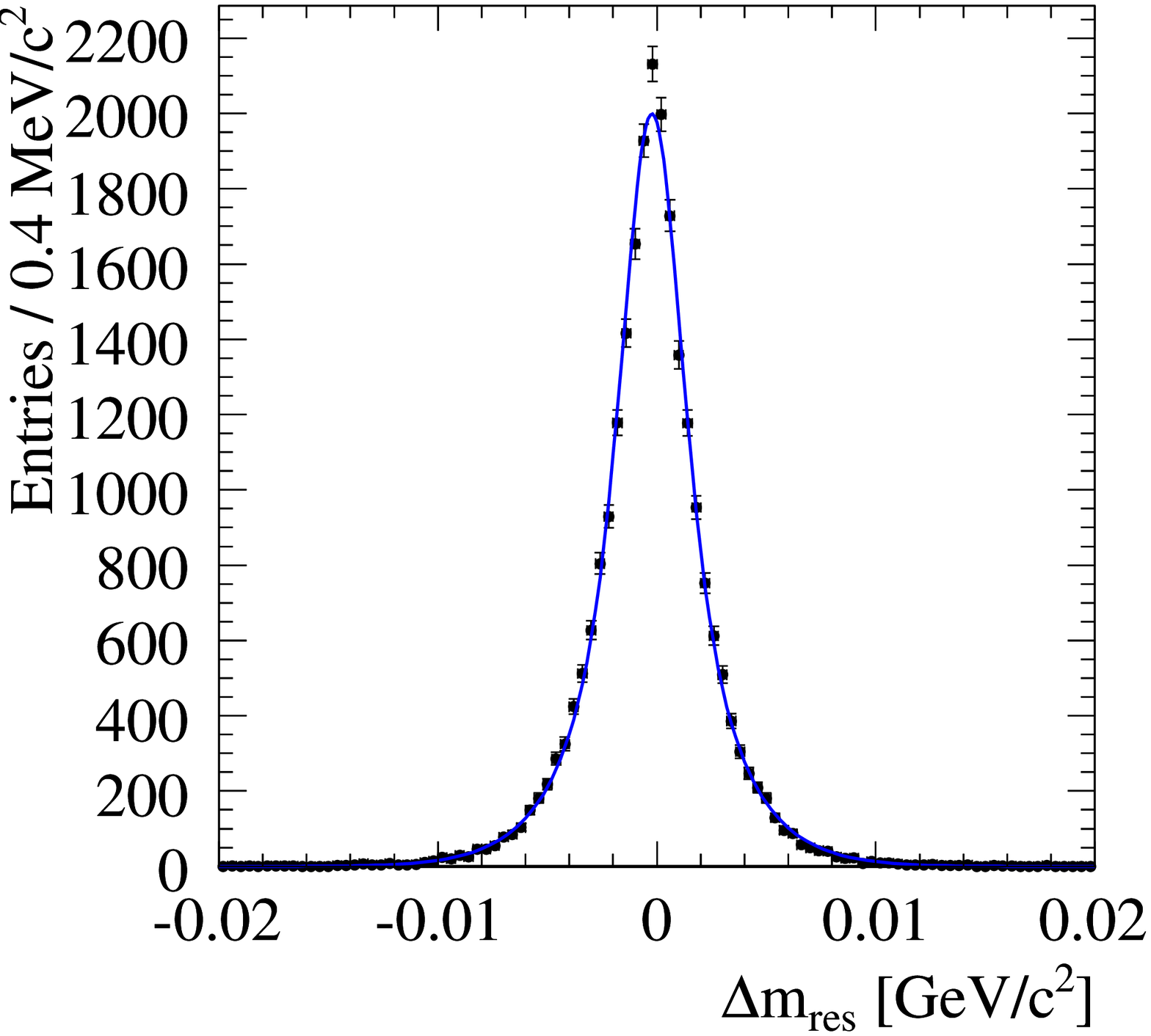}
\includegraphics[width=0.3\textwidth]{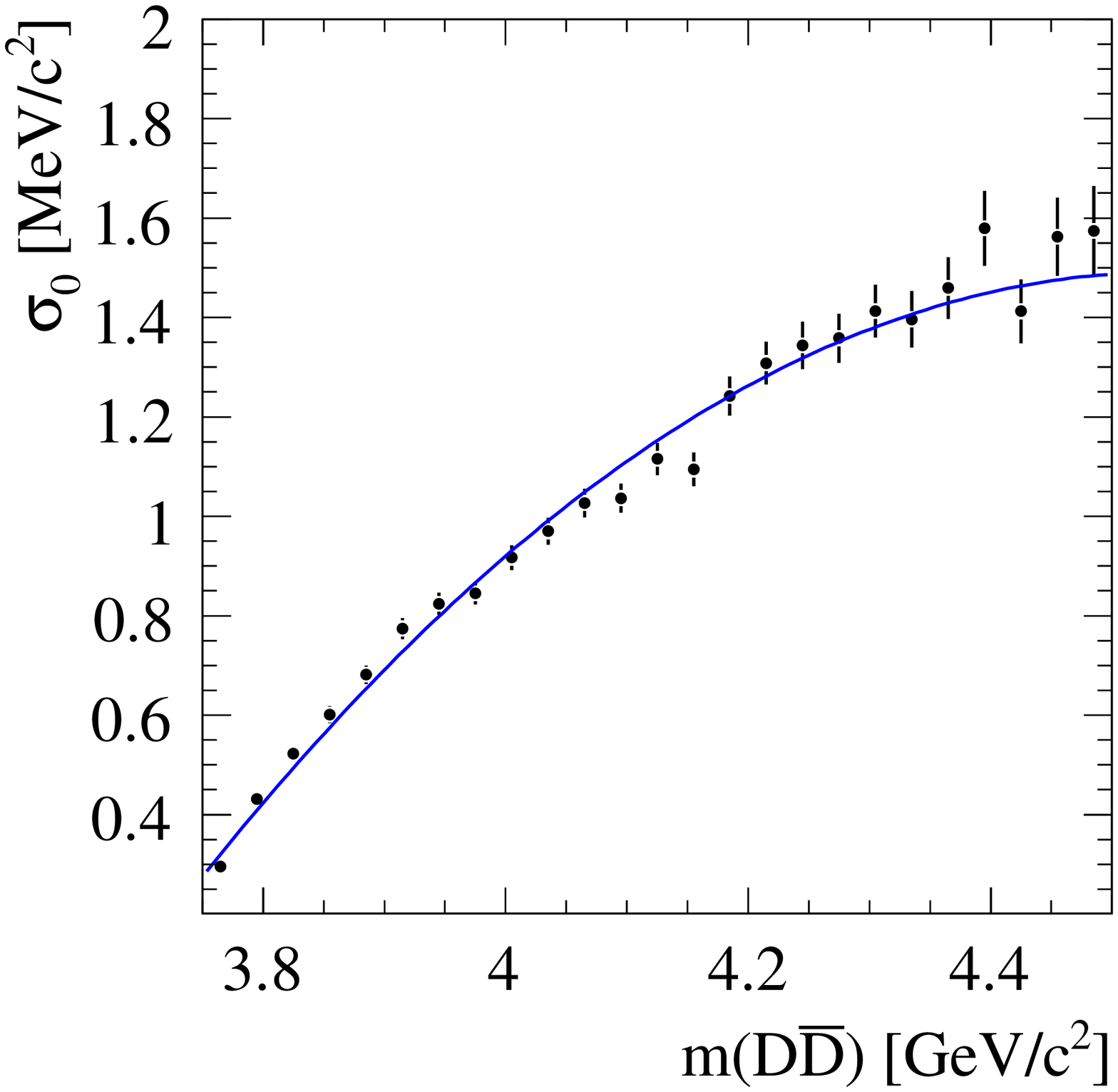}
\includegraphics[width=0.3\textwidth]{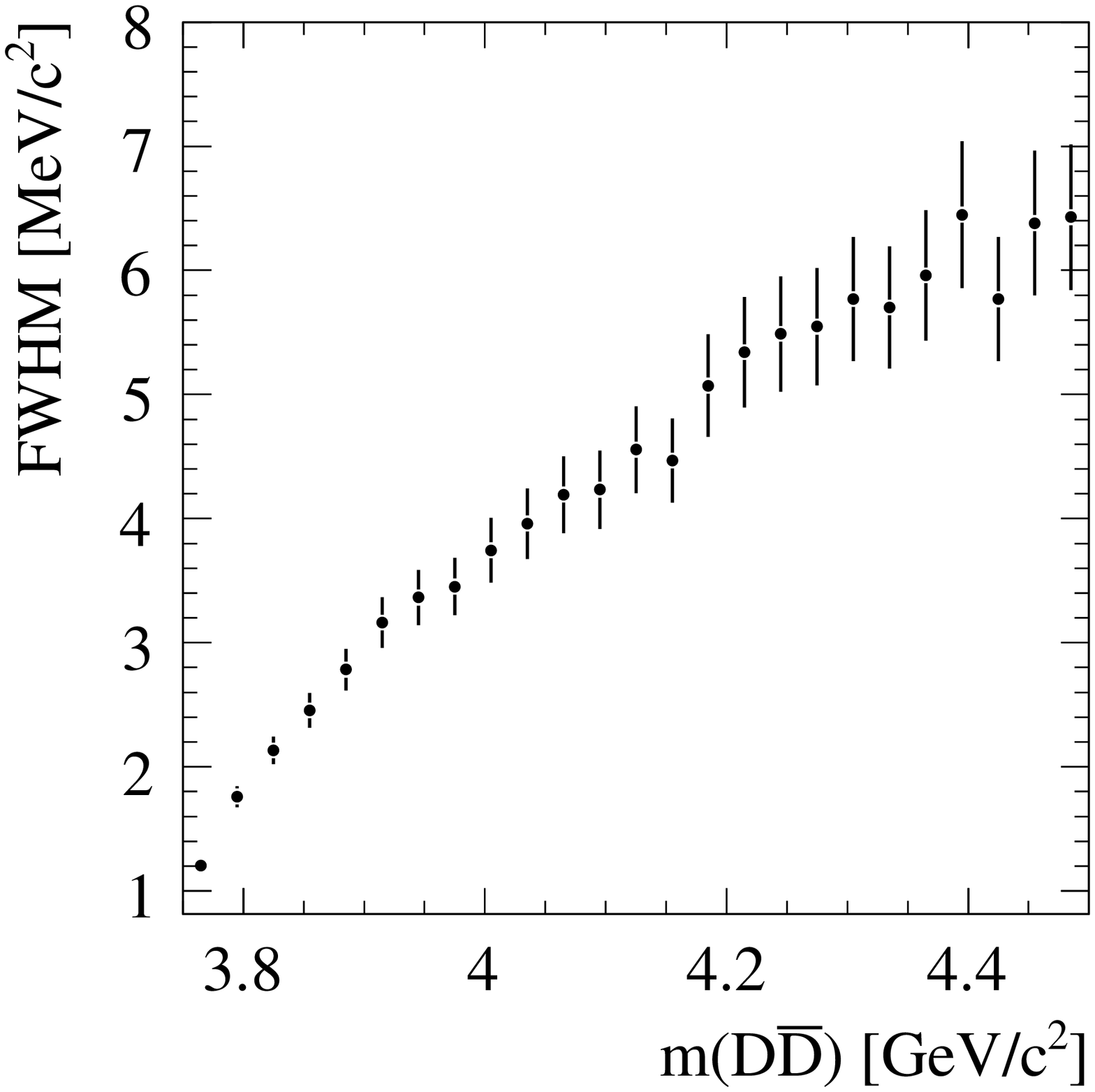}
\caption{(a) Detector resolution $\Delta m_{\rm res}$ for channel C6; the fitted curve is described in the text; (b) mass-dependence of the resolution function parameter $\sigma_{0}$; (c) mass-dependence of the FWHM of the resolution function.}
\begin{picture}(0,0)
\put(-6.5,5.7){(a)}
\put(-1.,5.7){(b)}
\put(4.5,5.7){(c)}
\end{picture}
\label{fig:reso}
\end{figure*}

\begin{figure}
\includegraphics[width=0.45\textwidth]{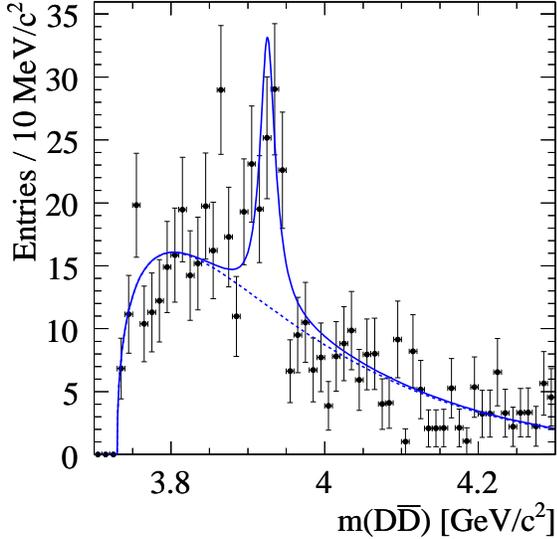}
\caption{Efficiency-corrected mean $D\Db$ mass distribution with standard fit. The dashed curve shows the background lineshape (see Sec.~\ref{sec_fit}).}
\label{fig:fit}
\end{figure}

Monte Carlo events are used for the calculation of the mass-dependent detector resolution. The mass resolution is determined by studying the difference between the reconstructed and the generated $D\Db$ mass ($\Delta m_{\rm res}$). As an example, the distribution for channel C6 is shown in Fig.~\ref{fig:reso}(a). A good description of the distribution is obtained using a multi-Gaussian fit (Eq.~(\ref{eq:mulg})). The parameters $r$ and $\sigma_{0}(m(D\Db))$ were determined for every decay channel. The variation of $\sigma_{0}(m(D\Db))$, which is parameterized by a second order polynomial, and of the width (FWHM) of the resolution function with increasing mass are shown in Figs.~\ref{fig:reso}(b) and~\ref{fig:reso}(c). For channel C6, $r = 5.380 \pm 0.137$ and $\sigma_{0}(m(D\Db)) = (-0.038 + 0.018m - 0.002m^{2})~\gevcc$, where $m(D\Db)$ is given in units of $\gevcc$. The distributions of Fig.~\ref{fig:reso} are well-described by the fitted curves shown. Comparing the generated $Z(3930)$ mass with the reconstructed MC value shows that the latter is systematically low by about $0.9\mevcc$, independently of the fit model. This effect is observed both in the combined fit and in fits to the individual channels. The measured \jpsi mass in the $\Kp\Km\pip\pim$ test sample (Sect.~\ref{sec_ggsel}) differs by the same value from the world average~\cite{Pd08}; this offset has been seen in other $\gamma\gamma$ studies at \babar~\cite{Au04} as well. Accordingly, the mass value obtained from the fit to data will be corrected by $+0.9\mevcc$. This offset value will also be used as a conservative estimate of the systematic uncertainty in the mass scale. The difference between the generated and reconstructed decay width values amounts to $0.14\mev$, and is discussed in Sec.~\ref{sec_syst} with respect to systematic error estimation.\\
\indent In order to describe the signal structure in data around $3.93\gevcc$ a relativistic Breit-Wigner function $BW(m)$ is used, where
\begin{equation}
\label{eq_relBWspin}
BW(m) = \left(\frac{p_{m}}{p_{m_{0}}}\right)^{2L+1}\left(\frac{m_{0}}{m}\right)\frac{F_{r}^{2}}{(m_{0}^{2}-m^{2})^{2}+\Gamma_{m}^{2}m_{0}^{2}}
\end{equation}
with $m_{0}$ as the nominal mass of the resonance; the Blatt-Weisskopf coefficients $F_{r}$ for different angular momentum values $L$ are given by
\begin{eqnarray}
F_{r}(L=0) & = & 1 \\
F_{r}(L=1) & = & \frac{\sqrt{1+(Rp_{m_{0}})^{2}}}{\sqrt{1+(Rp_{m})^{2}}}\\
F_{r}(L=2) & = & \frac{\sqrt{9+3(Rp_{m_{0}})^{2}+(Rp_{m_{0}})^{4}}}{\sqrt{9+3(Rp_{m})^{2}+(Rp_{m})^{4}}},
\end{eqnarray}
and the value
\begin{center}
$R=1.5~(\!\gevc)^{-1}$\\
\end{center}
is used, corresponding to the value given in Ref.~\cite{Hi72}. The mass-dependent width is given by
\begin{equation}
\Gamma_{m} = \Gamma_{r}\biggl(\frac{p_{m}}{p_{m_{0}}}\biggr)^{2L+1}\biggl(\frac{m_{0}}{m}\biggr)F_{r}^{2}
\end{equation}
with $\Gamma_{r}$ the total width of the resonance. Here the existence of other possible decay modes is ignored. The momentum of a given $D$ candidate in the $D\Db$ center of mass frame is denoted by $p_{m}$; $p_{m_{0}}$ is the corresponding value for $m = m_{0}$. In the standard fit, spin $J = 2$ ($L = 2$) is chosen on the basis of the angular distribution analysis described in Sec.~\ref{sec_ang}. \\
\indent The signal function is convolved with the mass- and decay-mode-dependent resolution model parameterized as discussed previously in this section. The background is parameterized by the function
\begin{equation}
D(m) \propto \sqrt{m^{2}-m_{t}^{2}}\left(m-m_{t}\right)^{\alpha}\exp\left[-\beta(m-m_{t})\right]
\label{eq_bibg}
\end{equation}
which takes the $D\Db$ threshold $m_{t}$ into account. In the lower mass region, the lineshape does not describe the background exactly. Other functional forms were tried (Sec.~\ref{sec_syst}), but no improvement was obtained. The data and the curves which result from the standard ($J=2$) fit are shown in Fig.~\ref{fig:fit}. \\
\indent From the unbinned maximum likelihood fit to the five mass spectra the $Z(3930)$ values $m_{0} = (3925.8 \pm 2.7)\mevcc$ and $\Gamma_{r} = (21.3 \pm 6.8)\mev$ are obtained for the mass and total width, respectively (all errors in this section are statistical only). The mass is corrected by $+0.9\mevcc$ as described above, resulting in a final mass value of $(3926.7 \pm 2.7)\mevcc$. The efficiency-corrected yield amounts to $N = (76 \pm 17)$ signal events. This value is based on weights around $1$ as discussed in Sec.~\ref{sec_effi}; taking the constant used to scale the efficiency into account (see Eq.~\ref{eq:effscale}), this corresponds to a total $Z(3930)$ signal of $N_{\epsilon^{B}} = (285 \pm 64)\times 10^{3}$ events. \\
\indent The statistical significance of the peak is $5.8\sigma$ and is derived from the difference $\Delta\ln{\cal L}$ between the negative logarithmic likelihood of the nominal fit and that of a fit where the parameter for the signal yield is fixed to zero. This is then used to evaluate a $p$-value:
\begin{equation}
p = \int_{2\Delta\ln{\cal L}}^{\infty}f(z;n_d) \,dz
\end{equation}
where $f(z;n_d)$ is the \chisq\ PDF and $n_d$ is the number of degrees of freedom, three in this case. We then determine the equivalent one-dimensional significance from this $p$-value.
\begin{figure}
\includegraphics[width=0.3\textwidth]{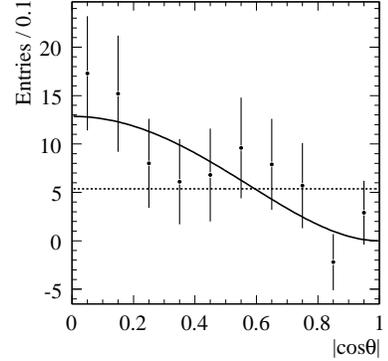}
\caption{Signal yield as a function of $\left|\cos\theta\right|$ derived from fits to the efficiency-corrected $D\Db$ spectrum. The solid curve is the expected distribution for spin 2 with dominating helicity-2 contribution, the dotted straight line is for spin 0.}
\label{fig:ang}
\end{figure}
\begin{figure}
\includegraphics[width=0.3\textwidth]{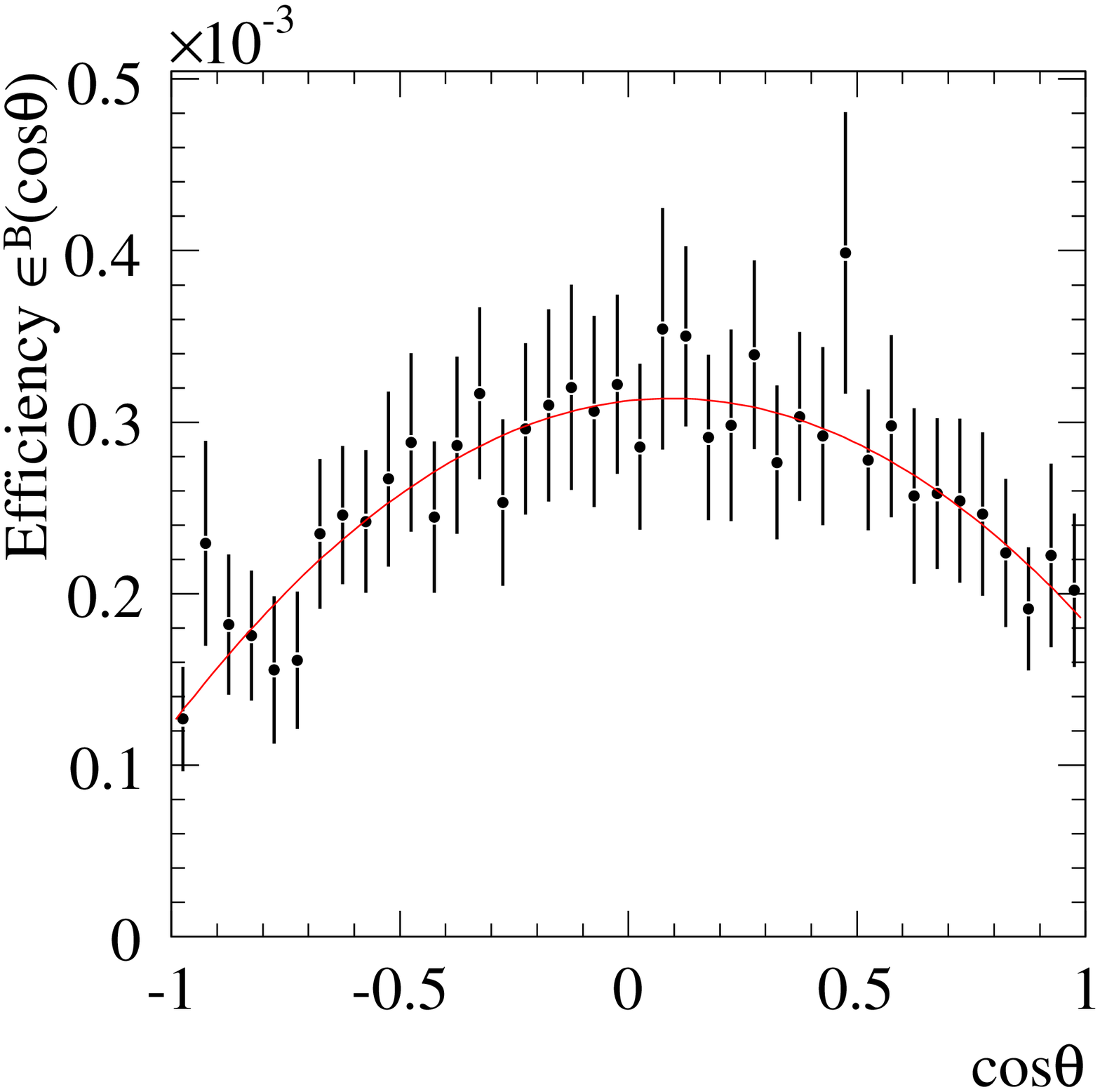}
\caption{Angular dependence of the weighted reconstruction efficiency $\epsilon^{B}(\cos\theta)$ based on Eq.~\ref{eq:meaneff}, described by a second order polynomial.}
\label{fig:angeff}
\end{figure}

\section{Angular distribution and spin of the {\boldmath $Z(3930)$} state}
\label{sec_ang}

General conservation laws limit the possibilities for the $J^{PC}$ values of the $Z(3930)$ state. For two-photon production the initial state has positive $C$-parity and hence the final state must have positive $C$-parity also. For the $D\Db$ final state, $C = (-1)^{L+S} = (-1)^{L}$ since the total spin $S$ is zero. Positive $C$-parity then implies that the $D\Db$ system must have orbital angular momentum $L$ which is even, and hence have even parity. It follows that for the $Z(3930)$ state $J^{PC} = J^{++}$ with $J = 0, 2, 4 \ldots$~In order to investigate the possible values of $J$, we have compared the decay angular distribution measured in the $Z(3930)$ signal region to the distributions expected for $J = 0$ and $J = 2$; higher spin values are very unlikely for a state only $200\mevcc$ above threshold.  \\
\indent The decay angle $\theta$ is defined as the angle of the $D$ meson in the $D\Db$ system relative to the $D\Db$ lab momentum vector. Figure~\ref{fig:ang} shows the $Z(3930)$ signal yield obtained from fits to the $D\bar{D}$ mass spectrum for ten regions of $\left|\cos\theta\right|$. The data have been weighted by a $\cos\theta$-dependent efficiency, which was determined in a similar manner as described in Sec.~\ref{sec_effi} for the mass-dependent efficiency (Fig.~\ref{fig:angeff}). In these fits, the mass and width of the resonance have been fixed to the values found in Sec.~\ref{sec_fit}, and Eq.~(\ref{eq_bibg}) has been used to describe the background. Other background models have been tried, obtaining distributions fully consistent with Fig.~\ref{fig:ang}.   \\
\indent The function describing the decay angular distribution for spin 2 has been calculated using the helicity formalism and has the form
\begin{equation}
\frac{dN}{d\cos\theta} \propto \sin^{4}\theta.
\label{eq_ggang2}
\end{equation}
It has been assumed that the dominating amplitude has helicity 2. This is in agreement with previous measurements~\cite{Ue08b} and theoretical predictions~\cite{Po86,Sc98}. The distribution of Eq.~(\ref{eq_ggang2}) was fitted to the experimental angular distribution, and a $\chi^{2}/{\rm NDF}$ value of 5.63/9 was obtained, with ${\rm NDF}$ indicating the number of degrees of freedom. For a flat distribution, which is expected for spin 0, a $\chi^{2}/{\rm NDF} = 15.55/9$ was obtained. It follows that the preferred $J^{PC}$ assignment is $2^{++}$.

\section{Two-photon width of the {\boldmath $Z(3930)$} state}
\label{sec_rawi}
\begin{figure}
   \centering
    \includegraphics[width=0.3\textwidth]{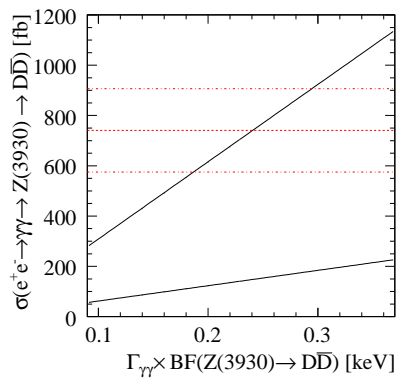}
\caption[]{Dependence of the cross-section $\sigma(\epem\to\gamma\gamma\to Z(3930) \to D\Db)$ on the two-photon width $\Gamma_{\gamma\gamma}\times {\cal B}(Z(3930)\to D\Db)$, calculated with the two-photon generator {\tt GamGam}. The upper solid line is for spin 2, while the lower solid line is for spin 0. The measured value (horizontal dashed line) and its uncertainty range (horizontal dot-dashed lines) are indicated.}
\label{fig_gg_wqgg}
\end{figure}

From the efficiency-corrected number of observed signal events, $N_{\epsilon^{B}}$, we determine the total experimental cross-section
\begin{eqnarray}
\label{eq_expcs}
\sigma_{\rm exp}(\epem \to \epem\gamma\gamma, \gamma\gamma & \to & Z(3930), Z(3930) \to D\Db) \nonumber \\ = N_{\epsilon^{B}}/\int \mathcal{L}dt & = & 741 \pm 166\fb
\end{eqnarray}
where the integrated luminosity for the data sample analyzed is $\int\mathcal{L}dt = (384 \pm 4)\invfb$ and the error is only statistical.\\
\indent On the other hand, the cross-section for $Z(3930)$ production is given by
\begin{equation}
\sigma(\epem\!\to\gamma\gamma, \gamma\gamma\!\to Z(3930)) = L\times F\times \sigma(\gamma\gamma\!\to Z(3930)) 
\end{equation}
with
\begin{eqnarray}
\sigma(\gamma\gamma\to Z(3930)) = \int 4\pi(2J+1)(\hbar c)^{2}10^{-6}m_{Z}^{3}\times \nonumber \\ \frac{\Gamma_{\gamma\gamma}}{\sqrt{K}m}\frac{\Gamma_{\rm tot}}{(m_{Z}^{2}-m^{2})^{2}+m_{Z}^{2}\Gamma_{\rm tot}^{2}}dm^{2}
\end{eqnarray}
and can be calculated using {\tt GamGam}. Here $L$ is the two-photon flux, $F$ is the form factor (see Sec.~\ref{sec_mcstud}), $m_{Z}$ ($\Gamma_{\rm tot}$) is the resonance mass (width), and $\Gamma_{\gamma\gamma}$ is the two-photon width of the resonance. The kinematical factor $K$ is given by $K = (q_{1}q_{2})^{2} - q_{1}^{2}q_{2}^{2}$ ($q_{i}$ represent four vectors of photons). Further information can be found in Refs.~\cite{Bo73}, \cite{Bu75} and \cite{Po86}. The cross-section depends on the spin of the resonance and on $\Gamma_{\gamma\gamma}$. It is plotted for $J = 2$ and $J = 0$ in Fig.~\ref{fig_gg_wqgg} as a function of $\Gamma_{\gamma\gamma}$. From a comparison to the experimental cross-section (Eq.~(\ref{eq_expcs})), the partial width $\Gamma_{\gamma\gamma}\times {\cal B}(Z(3930)\to D\Db)$ is found to have the value $(0.24 \pm 0.05)\kev$ when $J=2$ is chosen as the most probable spin value (see Sec.~\ref{sec_ang}).

\section{Systematic error estimation}
\label{sec_syst}
Several sources of systematic uncertainty have been considered for the mass, decay width, and signal yield of the $Z(3930)$ state. The yield determines the value of $\Gamma_{\gamma\gamma}\times {\cal B}(Z(3930)\to D\Db)$. The standard fit to the efficiency-corrected mass spectrum is repeated with appropriate modifications. The differences $\Delta$ between the results obtained and the standard results are used as estimates of systematic uncertainty. No correlations have been taken into account. The results are summarized in Table~\ref{tab_syst}. Deviations for the mass ($|\Delta m|$), total width ($|\Delta\Gamma|$) and two-photon width ($|\Delta(\Gamma_{\gamma\gamma}\times {\cal B})|$) are considered negligible if they are less than $0.05\mevcc, 0.05\mev$ and $0.0005\kev$, respectively. 

\subsection{Fit parameterization}
\label{ssec_systsignal}

\subparagraph{Signal Lineshape:}
The standard fit has assumed spin $J = 2$ for the resonance (Sec.~\ref{sec_fit}). Using different spin values and $R$ values has no significant impact on the results (Table~\ref{tab_syst}; $\Delta(\Gamma_{\gamma\gamma}\times {\cal B})$ numbers are given for spin $J = 2$ only).

\subparagraph{Background Description:}
Different parameterizations of the background in the $m(D\Db)$ distribution have been used. Besides the nominal background (Eq.~(\ref{eq_bibg})), the following background shape was tried
\begin{equation}
D^{\prime}(m) \propto \left(1 - \exp\left[\frac{-\left(m-\alpha\right)}{\beta}\right]\right)\left(\frac{m}{\alpha}\right)^{\beta}+\gamma\left(\frac{m}{\alpha}-1\right);
\label{eq_bdst}
\end{equation}
the fit had a slightly worse, but still acceptable, likelihood value. The mass value changes by $\Delta m = +0.4\mevcc$, the width by $\Delta\Gamma = +3.0\mev$, the signal yield by $+9$ events with respect to the standard fit, and $\Gamma_{\gamma\gamma}\times {\cal B}$ changes accordingly by $+0.029\kev$ (Table~\ref{tab_syst}). Other background models yield consistent estimates for this source of systematic uncertainty.

\subsection{Detector resolution}

\subparagraph{Fit Precision and Mass Scale:}
A fit of the convolution of signal lineshape and resolution model to the MC sample has been performed. The mass offset observed in MC has been included by correcting the mass value by $+0.9\mevcc$. As a conservative estimate, this number is also used as the systematic uncertainty for the mass scale. The deviation between the generated width and the value obtained from the fit is $0.14\mev$, and again this is used as a conservative estimate of systematic uncertainty. Based on the uncertainty of the width, a value of $\Delta\Gamma_{\gamma\gamma}\times {\cal B} = 0.001\kev$ is derived.

\subparagraph{Resolution Model:}
The parameters of the multi-Gaussian resolution model were modified. The number of steps was enlarged from 25 to 35, the total convolution range for each data point enlarged by $+0.02\mevcc$, and the parameter $r$ of the multi-Gaussian was varied within its fit uncertainty $\delta r$. The corresponding shifts in the mass are $\Delta m = +0.2$, $<0.05$ and $<0.05\mevcc$. For $\Delta\Gamma$, shifts of $-0.2$, $-0.9$ and $-0.1\mev$ are obtained; from the modified signal yield, shifts of $-0.003$, $-0.003$ and $<0.0005\kev$ were obtained for $\Gamma_{\gamma\gamma}\times {\cal B}$ (Table~\ref{tab_syst}).

\subsection{Combined reconstruction efficiency}

\subparagraph{Parameterization:}
The average mass-dependent reconstruction efficiency has been parameterized by a straight line in the standard fit (Fig.~\ref{fig:effi}). Using a fit with a second order polynomial, the width changes by $-0.4\mev$; no mass shift was observed with respect to the standard fit result. For the signal yield, $+1$ entry is obtained; this yields no significant shift for $\Gamma_{\gamma\gamma}\times {\cal B}$ (Table~\ref{tab_syst}). 

\subparagraph{Tracking and Neutrals Correction:}
For the tracking efficiency a correction by $-0.8\%$ is applied per charged-particle track. This gives a correction factor of $0.968$ for modes N4, N5, and $0.953$ for N6, N7 and C6. The systematic uncertainty assigned to the tracking efficiency is $1.4\%$ per track for decays with more than 5 charged particle tracks and $1.3\%$ otherwise. The resulting uncertainty for $\Gamma_{\gamma\gamma}\times {\cal B}$ is $0.022\kev$. Concerning efficiency corrections for neutral particles, a correction factor of $0.984$ with an uncertainty of $3 \%$ per $\piz$ is used for modes N5 and N7. The resulting uncertainty for $\Gamma_{\gamma\gamma}\times {\cal B}$ is $0.003\kev$ (Table~\ref{tab_syst}).

\subparagraph{Uncertainty on the {\boldmath $D$} Branching Fractions:}
The errors on the $D$ branching fractions have been taken into accout by varying the values of ${\cal B}_{i}$ used in Eq.~(\ref{eq:meaneff}) within their standard deviations. No significant change is observed in mass and decay width. For the two-photon width $\Delta(\Gamma_{\gamma\gamma}\times {\cal B}) = \pm0.010\kev$ is obtained (Table~\ref{tab_syst}).

\subparagraph{Effect of Angular Distribution on Efficiency:}
The MC data sample used to obtain the efficiency and resolution was generated with a flat distribution in $\cos\theta$. To estimate the effect of the angular distribution on the reconstruction efficiency, a MC sample described by a $\sin^{4}\theta$ distribution has been generated and reconstructed. Comparing these reconstructed data with the nominal MC sample, the mean efficiencies differ by $8\%$, relatively, resulting in $\Delta(\Gamma_{\gamma\gamma}\times {\cal B}) = \pm0.018\kev$.   
 
\begin{table*}
\caption{Results of the systematic uncertainty studies for the mass, decay width and efficiency-corrected signal yield of the $Z(3930)$ state. Listed are the differences with respect to the standard values. $\Delta(\Gamma_{\gamma\gamma}\times {\cal B})$ numbers are given for spin $J = 2$ only. For the combined error, the values are added in quadrature.}
\begin{ruledtabular}
  \begin{tabular}{lccc} 
    Source of Systematic Uncertainty & $\Delta m(Z(3930))$ & $\Delta\Gamma(Z(3930)) $ & $\Delta(\Gamma_{\gamma\gamma}\times {\cal B})$ \\
    & [$\!\mevcc$] & [$\!\mev$] & [$\!\kev$] \\ \hline 
    Choice of Spin $J=1, J=0$ & $<0.05$ & $<0.05$ & $-$  \\ 
    Value of $R$ (Breit-Wigner) & $<0.05$ & $<0.05$ & $<0.0005$  \\ 
    Background $D^{\prime}(m)$ & $0.4$ & $3.0$ & $0.029$ \\ 
    Fit precision and mass scale & $0.9$ & $0.1$ & $0.001$ \\ 
    Convolution steps = 35 & $0.2$ & $0.2$ & $0.003$  \\ 
    Convolution range $+0.02\mevcc$ & $<0.05$ & $0.9$ & $0.003$ \\ 
    Resolution multi-Gauss $r \pm \delta r$ & $<0.05$ & $0.1$ & $<0.0005$ \\ 
    Combined reconstr. efficiency: polynomial & $<0.05$ & $0.4$ & $<0.0005$ \\ 
    Tracking efficiency correction & $<0.05$ & $<0.05$ & $0.022$ \\ 
    $\piz$ efficiency correction  & $<0.05$ & $<0.05$ & $0.003$ \\ 
    Error in $D$ branching fractions & $<0.05$ & $<0.05$ & $0.010$ \\ 
    Efficiency: angular distribution & $-$ & $-$ & $0.018$ \\
    Generator precision & $-$ & $-$ & $0.007$ \\ 
    Choice of Form Factor & $-$ & $-$ & $0.002$ \\ 
    PID & $0.4$ & $1.8$ & $0.004$ \\ 
    Uncertainty in $D$ mass & $0.3$ & $-$ & $-$ \\ 
    Luminosity & $-$ & $-$ & $0.002$  \\  \hline
    Combined error & $\pm 1.1$ & $\pm 3.6$ & $\pm 0.04$ \\ 
\end{tabular}
\label{tab_syst}
\end{ruledtabular}
\end{table*}

\subsection{Cross-section calculation from GamGam}

\subparagraph{Precision:}
In Sec.~\ref{sec_mcstud} a relative uncertainty of $\pm3~\%$ was obtained for the calculated cross-section. Propagating this error into the calculation of $\Gamma_{\gamma\gamma}\times {\cal B}$, an uncertainty $\Delta(\Gamma_{\gamma\gamma}\times {\cal B}) = \pm 0.007\kev$ results. 

\subparagraph{Form Factor:}
In the standard analysis the form factor of Eq.~(\ref{eq_ff}) has been used with $m_{v} = m(\jpsi)$. In order to estimate potential systematic effects, the cross-section was evaluated using a model predicted by perturbative QCD~\cite{Fe97}
\begin{equation}
  F = \frac{1}{\left(1-q_{1}^{2}/m_{v}^{2}-q_{2}^{2}/m_{v}^{2}\right)^{2}}.
\label{eq_ffalt}
\end{equation}
The cross-section calculated with {\tt GamGam} does not increase significantly ($\approx 0.1\%$) compared to that obtained using Eq.~(\ref{eq_ff}). Simultaneously the experimental efficiency decreases by $1\%$, so that the net effect on $\Gamma_{\gamma\gamma}\times {\cal B}$ is small. Similar effects have been observed when data and calculations with and without $q^{2}$ selection criteria are compared~\cite{Ue96,Ue08,Ue09}, and also in a previous CLEO analysis~\cite{Do94}. As a result a systematic uncertainty of $\pm 1\%$ is attributed to form factor uncertainty and this yields a deviation $\Delta(\Gamma_{\gamma\gamma}\times {\cal B}) = \pm 0.002\kev$.

\subsection{Other uncertainties}
\label{ssec_systother}

\subparagraph{Particle Identification (PID):}
For PID studies, the pion selection criteria have been tightened significantly, and the efficiency has been recalculated accordingly. The fit to the mass spectrum yields a change of $-0.4\mevcc$ for the mass and $-1.8\mev$ for the width. For $\Delta(\Gamma_{\gamma\gamma}\times {\cal B})$ a change of $-0.004\kev$ results.

\subparagraph{{\boldmath $D$} Mass Uncertainty:}
The uncertainty of the $D$ meson mass is taken into account. Both for $\Dz$ and $D^{\pm}$, the uncertainty is $0.17\mevcc$~\cite{Pd08}, which results in an uncertainty of $\pm 0.34\mevcc$ in the mass of the $Z(3930)$ state.
   
\subparagraph{Integrated Luminosity Uncertainty:}
For the integrated luminosity, an uncertainty of $\pm 1\%$ is assigned. From this an uncertainty $\Delta(\Gamma_{\gamma\gamma}\times {\cal B}) = \pm 0.002\kev$ is obtained.

\subsection{Total systematic uncertainty}
The systematic uncertainty estimates discussed in Secs.~\ref{ssec_systsignal} -~\ref{ssec_systother} are summarized in Table~\ref{tab_syst}. The individual estimates are combined in quadrature to yield net systematic uncertainty estimates on the Z(3930) mass, total width and value of $\Gamma_{\gamma\gamma}\times {\cal B}(Z(3930)\to D\Db)$ of $1.1\mevcc$, $3.6\mev$ and $0.04\kev$, respectively, as reported on the last line of Table~\ref{tab_syst}.

\section{Summary}
\label{sec_summ}
In the $\gamma\gamma \to D\Db$ reaction a signal in the $D\Db$ mass spectrum has been observed near $3.93\gevcc$ with a significance of $5.8\sigma$ which agrees with the observation of the $Z(3930)$ resonance by the Belle Collaboration~\cite{Ue06}. The mass and total width of the $Z(3930)$ state are measured to be \mbox{$(3926.7 \pm 2.7 ({\rm stat}) \pm 1.1 ({\rm syst}))\mevcc$} and \mbox{$(21.3 \pm 6.8 ({\rm stat}) \pm 3.6 ({\rm syst}))\mev$}, respectively. \\
\indent The production and decay mechanisms allow only positive parity and $C$-parity, and an analysis of the $Z(3930)$ decay angular distribution favors a tensor over a scalar interpretation. The preferred assignment for spin and parity of the $Z(3930)$ state is therefore $J^{PC} = 2^{++}$. The product of the branching fraction to $D\Db$ times the two-photon width of the $Z(3930)$ state is measured to be $\Gamma_{\gamma\gamma}\times {\cal B}(Z(3930)\to D\Db) = (0.24 \pm 0.05 ({\rm stat}) \pm 0.04({\rm syst}))\kev$, assuming spin $J = 2$. The parameters obtained are consisten\t with the Belle results, and with the expectations for the $\chictwo(2P)$ state.  

\begin{acknowledgments}
We are grateful for the
extraordinary contributions of our \pep2\ colleagues in
achieving the excellent luminosity and machine conditions
that have made this work possible.
The success of this project also relies critically on the
expertise and dedication of the computing organizations that
support \babar.
The collaborating institutions wish to thank
SLAC for its support and the kind hospitality extended to them.
This work is supported by the
US Department of Energy
and National Science Foundation, the
Natural Sciences and Engineering Research Council (Canada),
the Commissariat \`a l'Energie Atomique and
Institut National de Physique Nucl\'eaire et de Physique des Particules
(France), the
Bundesministerium f\"ur Bildung und Forschung and
Deutsche Forschungsgemeinschaft
(Germany), the
Istituto Nazionale di Fisica Nucleare (Italy),
the Foundation for Fundamental Research on Matter (The Netherlands),
the Research Council of Norway, the
Ministry of Education and Science of the Russian Federation,
Ministerio de Educaci\'on y Ciencia (Spain), and the
Science and Technology Facilities Council (United Kingdom).
Individuals have received support from
the Marie-Curie IEF program (European Union) and
the A. P. Sloan Foundation.
\end{acknowledgments}

\end{document}